\documentclass[10pt,conference]{IEEEtran}
\usepackage{cite}
\usepackage{amsmath,amssymb,amsfonts}
\usepackage{algorithmic}
\usepackage{graphicx}
\usepackage{textcomp}
\usepackage{xcolor}
\usepackage[hyphens]{url}
\usepackage{makecell}
\usepackage{multirow}
\usepackage{pifont}
\newcounter{cnt}
\newcommand{\blackcir}[1]{\setcounter{cnt}{201}\addtocounter{cnt}{#1}\ding{\value{cnt}}}
\newcommand{\whitecir}[1]{\setcounter{cnt}{191}\addtocounter{cnt}{#1}\ding{\value{cnt}}}

\def\BibTeX{{\rm B\kern-.05em{\sc i\kern-.025em b}\kern-.08em
T\kern-.1667em\lower.7ex\hbox{E}\kern-.125emX}}

\newcommand{\archname}[1]{RePAST}
\newcommand{\revisionadd}[1]{#1}

\title{\archname{}: A ReRAM-based PIM Accelerator for Second-order Training of DNN}
\author{\IEEEauthorblockN{Yilong~Zhao$^1$,Li~Jiang$^1$,Mingyu~Gao$^2$,Naifeng~Jing$^1$,Chengyang~Gu$^1$,Qidong~Tang$^1$,\\Fangxin~Liu$^1$,Tao~Yang$^1$,Xiaoyao~Liang$^1$*}
\IEEEauthorblockA{$^1$School of Electronic Information and Electrical Engineering, Shanghai Jiaotong University, Shanghai, China\\$^2$ Institute for Interdisciplinary Information Sciences, Qinghua University, Beijing, China\\
sjtuzyl@sjtu.edu.cn~jiangli@cs.sjtu.edu.cn~gaomy@tsinghua.edu.cn~\\\{sjtuj,zeb19980914,tangqidong,liufangxin,yt594584152\}@sjtu.edu.cn~liang-xy@cs.sjtu.edu.cn}
}

\begin{document}
\maketitle
\thispagestyle{plain}
\pagestyle{plain}


\begin{abstract}

The second-order training methods can converge much faster than first-order optimizers in DNN training.
  This is because the second-order training utilizes the inversion of the second-order information (SOI) matrix to find a more accurate descent direction and step size.
  However, the huge SOI matrices bring significant computational and memory overheads in the traditional architectures like GPU and CPU.
On the other side, the ReRAM-based process-in-memory (PIM) technology is suitable for the second-order training because of the following three reasons:
First, PIM's computation happens in memory, which reduces data movement overheads;
  Second, ReRAM crossbars can compute SOI's inversion in $O\left(1\right)$ time;
Third, if architected properly, ReRAM crossbars can perform matrix inversion and vector-matrix multiplications which are important to the second-order training algorithms.

  Nevertheless, current ReRAM-based PIM techniques still face a key challenge for accelerating the second-order training.
  The existing ReRAM-based matrix inversion circuitry can only support 8-bit accuracy matrix inversion and the computational precision is not sufficient for the second-order training that needs at least 16-bit accurate matrix inversion.
  In this work, we propose a method to achieve high-precision matrix inversion based on a proven 8-bit matrix inversion (INV) circuitry and vector-matrix multiplication (VMM) circuitry.
  We design \archname{}, a ReRAM-based PIM accelerator architecture for the second-order training.
  Moreover, we propose a software mapping scheme for \archname{} to further optimize the performance by fusing VMM and INV crossbar.
Experiment shows that \archname{} can achieve an average of 115.8$\times$/11.4$\times$ speedup and 41.9$\times$/12.8$\times$energy saving compared to a GPU counterpart and PipeLayer on large-scale DNNs.

\end{abstract}

\section{Introduction}
\label{sec:introduction}
Most prevalent optimizers for neural network training, including Stochastic Gradient Descent (SGD) \cite{momentumQIAN1999145}, Adagrad \cite{adagrad10.5555/1953048.2021068} and Adam \cite{kingma2014adam}, only use the information of the first-order gradient.
However, as the complexity of the neural network model increases, these optimizers take much more time to train the neural network.
For example, it takes over 29 hours to train a ResNet-50 on an 8-GPU cluster \cite{resnethe2015deep}.
To address the problem, second-order training algorithms are proposed to accelerate the training process \cite{kfacpmlr-v37-martens15,spngdosawa2020scalable,hessianfreekiros2013training}.
These algorithms take advantage of the inverse of the second-order information (SOI) matrix to accelerate the convergence.
For the common second-order optimization methods, e.g. Newton method and natural gradient method, their SOI matrices are Hessian Matrix (HM) and Fisher Information Matrix (FIM), respectively.
For small neural networks on small datasets, such as autoencoder on MINST, second-order training can reduce the iteration number by over 100$\times$\cite{kfacpmlr-v37-martens15}.
One major challenge of directly performing the two algorithms on large-scale DNNs is that the SOI size increases quadratically as the parameter number increases.
For example, a ResNet-50 network has $2.5\times10^7$ parameters, and the size of its HM will be around $6.3\times10^{14}$.
The large size of SOI brings the following two problems:
First, computing with SOI requires a significant amount of data movement.
Second, the complexity of matrix inversion on SOI is $O(n^3)$, which causes a prohibitive computational cost.
As a result, directly applying the second-order algorithms to DNNs in reality is actually much slower than the first-order algorithms.

Current second-order training algorithms approximate the SOI into several smaller matrices to reduce the overhead brought by SOI.
They treat the elements across different layers as 0.
K-FAC algorithm uses Kronecker decomposition to decompose the FIM of each layer into two smaller matrices \cite{kfacpmlr-v37-martens15}.
However, the size of SOI is still large.
For example, after this approximation, the size of SOI is still about $1.4\times10^8$ for ResNet-50.
ADAHESSIAN directly approximates the SOI into a diagonal matrix so that we do not need to invert the SOI matrix \cite{yao2021adahessian}.
However, this results in a large computational error such that little improvement in the convergence rate for the training process is observed.
Some algorithms trade off the overhead of SOI and the convergence rate, by approximating the SOI matrix into a block-diagonal matrix \cite{thorChen_Gao_Liu_Wang_Ni_Zhang_Chen_Ding_Huang_Wang_Wang_Yu_Zhao_Xu_2021,distributed8850515}.
However, limited by the GPU performance, the optimal block size is usually small.
For example, the optimal block size of THOR algorithm is only 128 \cite{thorChen_Gao_Liu_Wang_Ni_Zhang_Chen_Ding_Huang_Wang_Wang_Yu_Zhao_Xu_2021}.
With smaller block sizes, it takes more epochs to train due to higher approximate errors, causing longer convergence time, offsetting the algorithmic advantage of the second-order training.

ReRAM-based PIM is an emerging technique whose computation happens in the memory and has the inherent ability to accelerate vector-matrix multiplication (VMM) \cite{dot7544263}.
While each ReRAM cell can only differentiate a limited number of bit levels, which means the ReRAM crossbar itself can only perform low-precision VMM computation, high-precision VMM can be carried out by spliting it into multiple low-precision VMMs with bit-slicing scheme \cite{ISAAC}.
Based on this, Song et al. propose Pipelayer, a ReRAM-based architecture to accelerate the first-order training of DNN \cite{pipelayer7920854}.
The ReRAM crossbar can also accelerate the computation of matrix inversion (INV), such as the design proposed by Sun et al. \cite{invcircuit8914709}.
The time complexity of their circuit is $O\left(1\right)$. But the precision is still limited by the number of bits that the ReRAM cell can support.
The authors in \cite{predictioner9407108} propose an analog bit-slicing scheme and increase the precision of INV circuit to 8-bit.
They use analog amplifiers and combine the analog signals from bitlines into a more precise analog signal.
However, the method can not further improve the precision of the INV circuit because the noise level has surpassed the limit of the precision in the analog signals.

The characteristic of ReRAM-based PIM technology makes it suitable for accelerating the second-order training algorithms.
PIM can reduce the data movement of the large SOI and PIM can perform the matrix inversion in $O\left(1\right)$ time regardless of the matrix size, reducing the computational overhead of matrix inversion. Also, ReRAM-based crossbars can perform high-precision VMM through the bit-slicing technique.
However, the key challenge in designing PIM accelerators for the second-order training algorithms is that the current PIM matrix inversion circuit can only support up to 8-bit precision. This cannot meet the criterion for the second-order training algorithms, most of which require 16-bit precision or higher \cite{kfaccnnpmlr-v48-grosse16,kfacpmlr-v37-martens15,thorChen_Gao_Liu_Wang_Ni_Zhang_Chen_Ding_Huang_Wang_Wang_Yu_Zhao_Xu_2021,distributed8850515}.

In this work, we address this challenge and propose \archname{}, a ReRAM-based PIM accelerator for the second-order training.
We make the following contributions:

\begin{itemize}
      \item We propose a novel scheme to implement the high-precision matrix inversion based on multiple low-precision matrix inversion circuit.
            Leveraging Taylor Expansion, a high-precision matrix inversion can be decomposed into a series of low-precision matrix inversions chained with a VMM operation, both of which can be implemented with ReRAM-based crossbars.
      \item We propose \archname{}, a ReRAM-based accelerator architecture for the second-order training.
The \archname{} has a series delicate architecture design and mapping strategy to fuse INV and VMM operations smoothly that can support all the typical operations in the second-order training algorithms, including VMM and high-precision INV.
      \item 
            The mapping strategy can also reduce the overhead of SOI matrix.
            We find that some patterns in the dataflow graph vary in computation latency or memory footprint if we choose different mapping strategies.
            For each pattern, we propose a method to select the optimized mapping strategy for different layers.

\end{itemize}

The remainder of the paper is organized as follows:
Section \ref{sec:background} introduce the background of the second-order training algorithms and ReRAM-based VMM and low-precision INV circuits.
In Section \ref{sec:matinv}, we introduce a novel method for the high-precision matrix inversion and implement it based on the ReRAM-based VMM and low-precision INV circuits.
Section \ref{sec:architecture} presents the overview and the hardware design of the \archname{} architecture.
In Section \ref{sec:mapping}, the mapping scheme is explained.
Section \ref{sec:evaluation} presents the evaluation and analysis of \archname{}.
Section \ref{sec:related} discusses other related works.
Section \ref{sec:conclusion} concludes the paper.

\section{Background}
\label{sec:background}
\subsection{Second-Order Training}
\label{sec:background:so_algorithms}

First-order training algorithms only utilize the first-order information to obtain the direction and step for optimization.
The rule for parameter update is:
\begin{eqnarray}
      \theta & = & \theta - \eta \cdot \nabla_\theta J\left(\theta\right)
      \label{eqn:sgd_update}
\end{eqnarray}
where $\theta$ is the parameter of neural network, $J$ is the loss function, and $\eta$ is the learning rate.

Second-order training methods take advantage of the SOI matrix $H$  to obtain a better descent direction and step.
Since the SOI matrix depicts the surrounding landscape, the descent path is closer to the optimal descent path and can avoid getting stuck in the saddle point.
As a result, the convergence is faster and the number of iterations and epochs can be reduced.
For example, the second-order optimizer proposed in \cite{spngdosawa2020scalable} only needs half of the iterations or epochs to train a ResNet-50 compared to the first-order one.
The rule of parameter update in the second-order training is:
\begin{eqnarray}
      \theta & = & \theta - \eta \cdot H^{-1} \nabla_\theta J\left(\theta\right)
      \label{eqn:so_update}
\end{eqnarray}
where $H$ is derived from $J$, and $\theta$, decided by the definition of SOI.
However, this formulation cannot be directly implemented for DNN.
This is because the size of SOI scales quadratically with the number of parameters, which results in unacceptably large overheads for both computation and storage.
Therefore, most algorithms decouple the full SOI matrix by layers, regarding the second-order gradient between layers as 0.
We introduce two popular approaches below:

\textit{1) K-FAC Algorithm.}
K-FAC is derived from Natural Gradient method (NG), whose SOI is the Fisher information matrix (FIM) $\mathcal{F}$ \cite{kfacpmlr-v37-martens15}.
K-FAC decomposes each layer's FIM block into the Kronecker product of two small matrices $A$ and $G$, which are much smaller than the original FIM.
After the approximation and decomposition, the storage of the ResNet-50 SOI becomes 140 million, still large but acceptable for GPU.
Moreover, the Kronecker product can be transformed into matrix multiplication.
Therefore the parameter update rule becomes:
\begin{eqnarray}
      \theta & = & \theta - \eta \cdot A^{-1} \nabla_\theta J\left(\theta\right) G^{-1}
      \label{eqn:k-fac_update}
\end{eqnarray}
The matrix $A$ and $G$ are obtained by $A=a\cdot a^T$ and $G=g\cdot g^T$, where $a$ is the input feature map and $g$ is the gradient of the layer.
Therefore, the two SOIs can be obtained during the backpropagation efficiently.
In the fully connection layer (FC), the input feature map $a$ and gradient $g$ are directly multiplied with themselves when computing $A$ and $G$.
While for convolution layer, $a$ and $g$ are firstly reshaped to $c_{in}k^2\times hw$ and $c_{out}\times hw$ respectively, and then multiplied with themselves, where $c_{in}$, $c_{out}$, $k$, $h$ and $w$ are input/output channel, kernel size, height and width of the feature map, respectively.
Therefore, for convolution layer, $A\in \mathbb{R}^{c_{in}k^2\times c_{in}k^2}$ and $G\in \mathbb{R}^{c_{out}\times c_{out}}$ \cite{spngdosawa2020scalable}.

\textit{2) Gauss-Newton Algorithm.}
Gauss-Newton method is derived from the Newton method, whose SOI is the Hessian matrix (HM) $\mathcal{H}$ \cite{hessianfreekiros2013training,distributed8850515}.
Each layer's HM block is approximated by $H\approx \nabla_\theta J\left(\theta\right) B  \nabla_\theta J\left(\theta\right)^T $, where the matrix $B$ is decided by the loss function.
For example, when using cross entropy loss, $B$ becomes the identity matrix.

\begin{figure}
\includegraphics[width=0.9\linewidth]{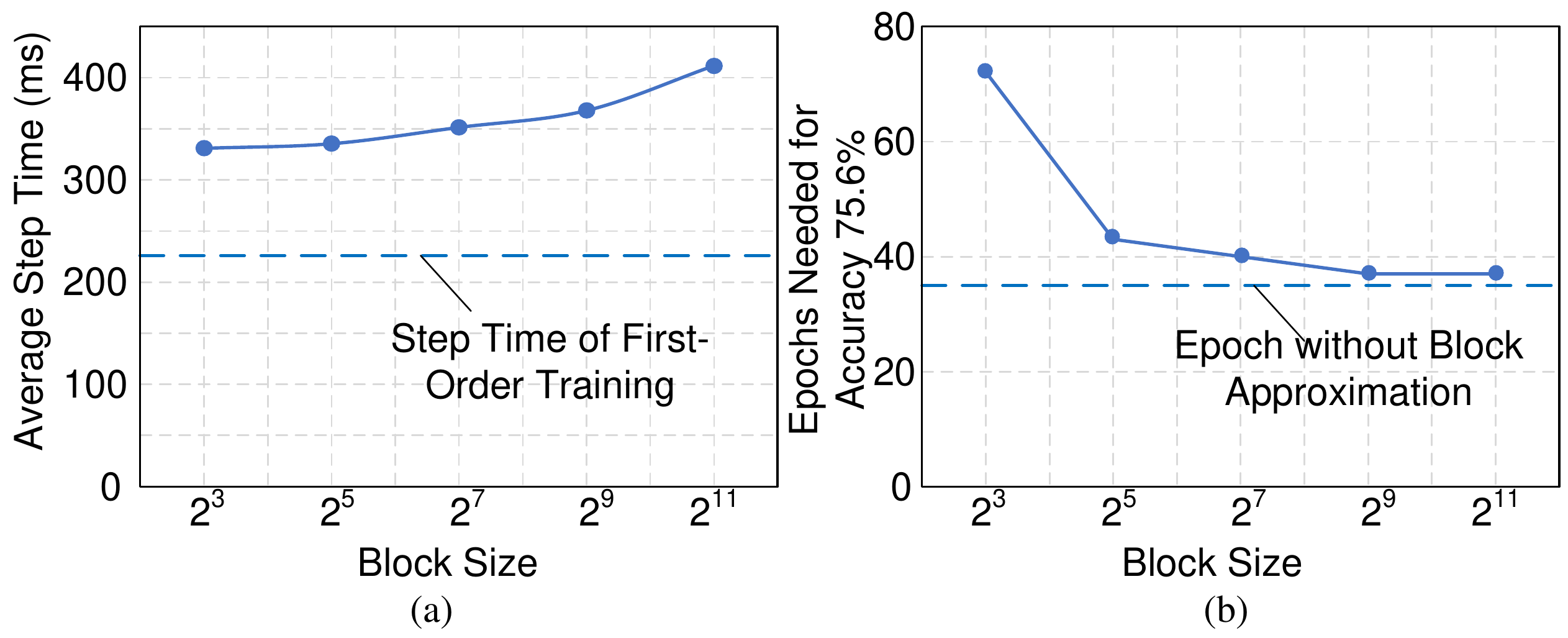}
\vspace{-10pt}
      \caption{ With different block sizes, (a) the step time of training ResNet-50 on GPU, and (b) the epoch number of training ResNet-50 to 75.6\% accuracy.
            The epoch number of ResNet-50 training without block approximation is 34~\cite{spngdosawa2020scalable}.}
      \label{fig:observation}
\vspace{-10pt}
\end{figure}

\textbf{Diagonal block approximation of SOI matrices.}
Nevertheless, even with the approximation in both algorithms, the computation of SOI is still a large overhead.
Therefore, some works further approximate FIM and HM into block-diagonal matrices \cite{thorChen_Gao_Liu_Wang_Ni_Zhang_Chen_Ding_Huang_Wang_Wang_Yu_Zhao_Xu_2021,distributed8850515} or even diagonal matrices \cite{yao2021adahessian}.
Therefore, the inversion of SOI matrix can be obtained by inverting each blocks, and reduce the computational overhead.
Take one convolution layer of ResNet-50's conv5\_x as an example.
The layer has 512 output channels, therefore, the matrix $G$ is $512\times512$ according to \cite{spngdosawa2020scalable}.
If we approximate $G$ into four $128\times128$ diagonal blocks, the storage is reduced by $4\times$.
Moreover, as the computational complexity of matrix inversion is $O\left(n^3\right)$, after approximation, the computing is reduced by $64\times$.
\figurename{}~\ref{fig:observation}(a) plot the step time of ResNet-50 training with different block sizes.
Using a smaller block size can reduce the execution time of SOI.
However, a small block size also means that more information in SOI has been discarded and the descent direction and step size may not be accurate.
This will actually slow down the overall convergence process in training because it requires more training epochs to converge.
As \figurename{}~\ref{fig:observation}(b) shows, the required epoch number of ResNet-50 training increases sharply with a shrinking block size.
Therefore, this approximation introduces a trade-off between the convergence rate and block size.
In \cite{thorChen_Gao_Liu_Wang_Ni_Zhang_Chen_Ding_Huang_Wang_Wang_Yu_Zhao_Xu_2021}, the result shows that a block size of $128$ should be suitable for a GPU. However, such a block size may slow down the convergence by 29.4\% and diminish the advantage of the second-order training.

\subsection{PIM-Based VMM and Matrix Inversion}

\begin{figure}
      \includegraphics[width=\linewidth]{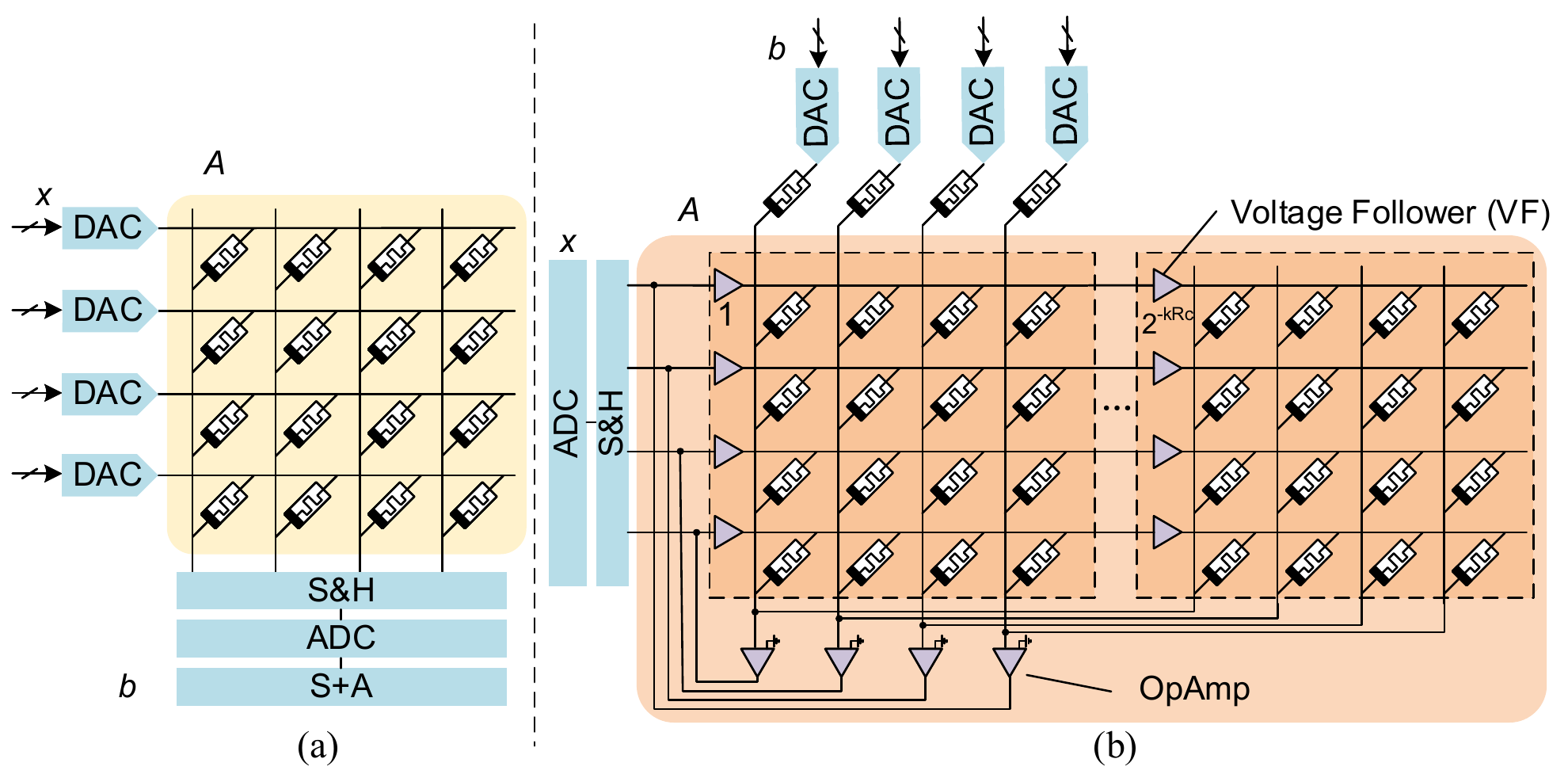}
\vspace{-20pt}
      \caption{ReRAM crossbar circuits for accelerating (a) vector-matrix multiplication (VMM) and (b) matrix inversion (INV).}
      \label{fig:vmm_inv}
\vspace{-10pt}
\end{figure}

Current works show that ReRAM-based crossbars offer great efficiency in computing vector-matrix multiplication (VMM) with PIM technologies \cite{dot7544263}.
One such ReRAM-based VMM accelerator is illustrated in \figurename{}~\ref{fig:vmm_inv}(a).
The matrix element values are programmed (i.e. written) into the crossbar, and the input vector is applied to the wordline as voltage levels through DACs.
According to the Kirchhoff's Law, the aggregated currents on the bitlines, after passing through ADCs, become the result of the VMM.
DNN inference and training require sufficient precision for VMM operations, e.g., 8-16 bits.
However, the state-of-the-art ReRAM cell can only handle 2-6 bits \cite{reramreviewRN21}.
Therefore, each element of the matrix is split by bits and written to several adjacent cells.
Moreover, the DAC's complexity is exponential to the resolution.
To allow for low-bit DACs, each input vector element is also split by bits and applied to a wordline across several cycles.
The intermediate results across bitlines and cycles are combined into the final result by shift-and-add operation (S+A).
This is the well-known \emph{bit-slicing} scheme to solve the precision problem in VMM~\cite{ISAAC, prime10.1109/ISCA.2016.13}.

Recent works also use ReRAM crossbars to accelerate matrix inversion (INV) with the help of extra analog components~\cite{invcircuit8914709,predictioner9407108,invcombine10.1145/3453688.3461510}.
The circuit is shown in \figurename{}~\ref{fig:vmm_inv}(b).
Suppose a matrix $A$ is $Q_A$-bit quantitized, and a ReRAM cell has $R_c$ bits.
$A$ is bit-sliced and each slice is programmed to a separate ReRAM crossbar.
The $i^{\mathrm{th}}$ ReRAM crossbar's voltage followers' (VFs') gain is set to $2^{-iR_c}$.
A vector $-b$ is applied as voltage levels on the bitline resistances of the crossbar.
The currents on the bitlines of the crossbars are aggregated, and are fed back with open-loop operational amplifiers (OpAmps) to the wordlines.
According to the virtual short and virtual open characteristic of the OpAmps, the voltages on the wordlines would converge to the values that make both the current and voltage on the bitlines to 0.
Then the wordline voltage $x$, $A$ and the bitline voltage $b$ will satisfy Eqn. 4, effectively achieving a matrix INV in Eqn. 5:
\begin{eqnarray}
      x \cdot A & = & b \\
      x & = & A^{-1} \cdot b
      \label{eqn:matrix-inv}
\end{eqnarray}
The convergence time is decided by the dynamic characteristic of the OpAmps.
If the OpAmp has a wider bandwidth, the convergence is faster.
If we can keep the convergence time within one cycle, the circuit can compute the matrix inversion in $O(1)$ time.
For example, the circuit in \cite{predictioner9407108} can converge within 50ns, much shorter than the cycle time for PIM architectures (typically around 100ns~\cite{ISAAC}).



\begin{figure}
    \centering
\includegraphics[width=0.7\linewidth]{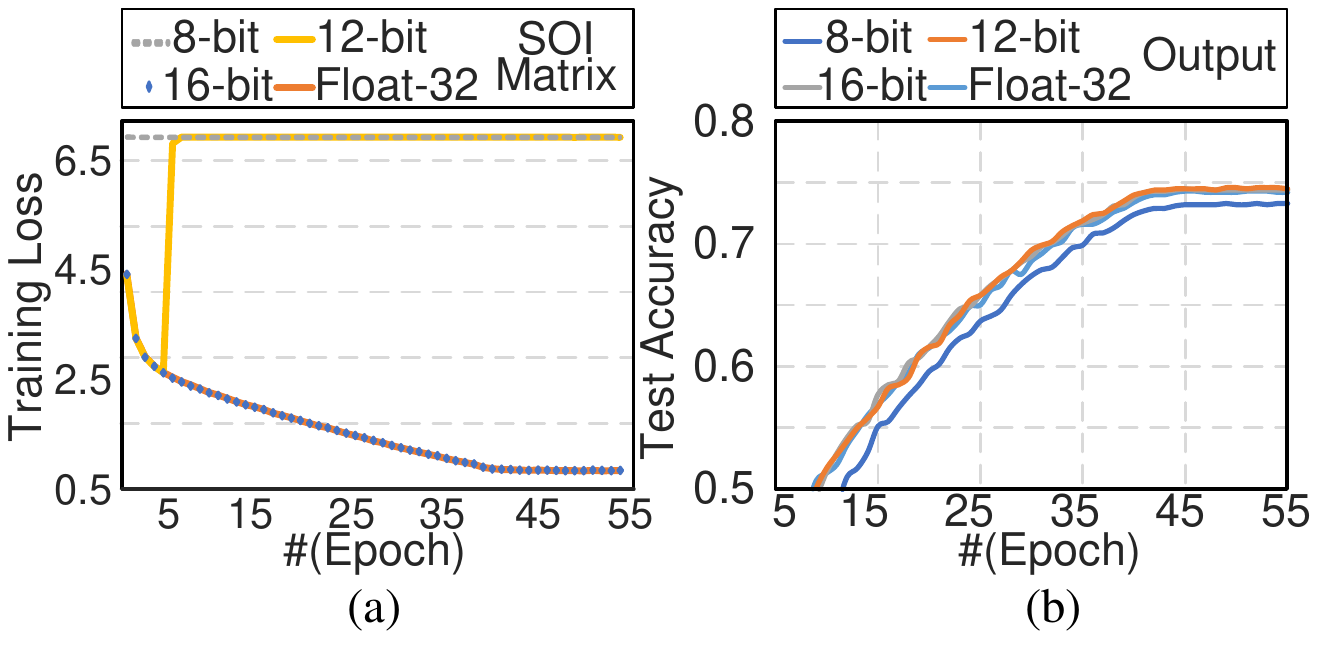}
\vspace{-10pt}
\caption{(a) Loss curve when training ResNet-50 with 8-bit, 12-bit, 16-bit and 32-bit (full-precision) second-order information (SOI) matrices. (b) The test accuracy in ResNet-50 with different quantization levels to the result of matrix inversion, which shows that 16-bit result is sufficient for second-order training.}
    \label{fig:bit_slice_observation}
\vspace{-15pt}
\end{figure}

Although such ReRAM-based INV circuitry has been recently proposed to compute matrix inversion, its precision is still quite limited and cannot meet the requirement of DNN training.
\figurename{}~\ref{fig:bit_slice_observation}(a) depicts the training loss curve when training ResNet-50 \cite{resnethe2015deep} with 8-bit, 12-bit, 16-bit and full-precision SOI matrix using the same learning rate.
Training with 8-bit and 12-bit SOI matrix results in non-convergence.
This is because the SOI matrix's elements are small values, and quantization errors may lead to large deviations in the matrix inversion results.
The requirement of SOI matrix precision is at least 16-bit.
This precision exceeds the limit of existing matrix inversion circuitry
~\cite{invcircuit8914709,predictioner9407108,invcombine10.1145/3453688.3461510}.
\figurename{}~\ref{fig:bit_slice_observation}(b) depicts the test accuracy with different quantization levels to the result of matrix inversion.
A less accurate matrix inversion result also degrades the training effect.
At least 16-bit accurate matrix inversion result is required.
Meanwhile, the precision of $x$ and $b$ is limited by ADC and DAC resolution, respectively.
Although the matrix $A$ is programmed to multiple crossbars, the precision of $A$ can be heavily limited by the noise on the bitlines.
For example, \cite{predictioner9407108} only supports a matrix inversion for 8-bit accuracy, with a 6-bit input $b$ and an 8-bit output $x$, and can only be used as one preconditioner.
Note that unlike VMM, the INV circuit does not allow a bit-slicing scheme, because matrix inversion does not have a distributive law.

\section{High-Precision Matrix Inversion}
\label{sec:matinv}


\subsection{The Algorithm for High-Precision Matrix Inversion}
\label{sec:matinv:scheme}

Given a matrix $A$ and a vector $b$, matrix inversion aims to solve $x=A^{-1}\cdot b$.
We denote the numbers of bits for the three quantized values as $Q_A$, $Q_b$, and $Q_x$, respectively.
Second-order training usually requires high-precision quantizations, e.g., $Q_A=16$, $Q_b=16$, and $Q_x=16$.
The resolution of DAC, ADC, and the number of bits a ReRAM cell can support is denoted as $R_{DAC}$, $R_{ADC}$ and $R_{c}$.
We connect $n$ crossbars together by VFs as \cite{predictioner9407108} does as shown in \figurename{}~\ref{fig:vmm_inv}(b).
Hence, the $k$ crossbars can work together for INV of a $(k\cdot R_{c})$-bit matrix $A$.
However, due to the noise impact in analog circuits, constructing a 16-bit matrix $A$ is not realistic.
Moreover, the resolutions of DAC and ADC still limit the precision of $x$ and $b$.
To address these issues, we propose a novel scheme for the INV circuit in this subsection.
The scheme can be divided into 3 steps, for $b$, $x$, and $A$, respectively:

\textit{1) Bit-slicing scheme for input $b$.}
For vector $b$, we can slice it into  $N_b=\left\lceil Q_{b}/R_{DAC}\right\rceil$ slices, with each slice having $R_{DAC}$ bits.
Thus, $b = \sum_{i=0}^{N_b-1} b_i\cdot2^{i\cdot R_{DAC}}$, where $b_i$ is the $i^{\mathrm{th}}$ slice of $b$.
We have:
\begin{eqnarray}
    x & = & A^{-1} \cdot b = \sum_{i=0}^{N_b-1} \left(A^{-1}\cdot b_i\right)\cdot2^{i\cdot R_{DAC}}
\label{eqn:b-slice}
\end{eqnarray}
Therefore, we can first calculate the matrix inversion on each slice of $b$ and then ``shift-and-add'' (S+A) the partial results to obtain the final result.

\textit{2) Bit-slicing scheme for output $x$.}
For the output vector $x$, the analog signals converged on the bitlines actually contain the complete information.
But due to ADCs' limited resolution, only the first $R_{ADC}$ bits can be quantized to digital values.
Therefore, we repeat the following steps iteratively until all the bits are quantized by the ADCs.
In the $j^{\mathrm{th}}$ iteration, one vector $b_{j}$ is applied to DAC and the result is quantized by the ADCs as $x_{j}$.
We will then compute $b_{j+1}=\left(b_{j}-A\cdot x_{j}\right)\cdot2^{R_{ADC}}$.
After that, we repeat the loop, applying $b_{j+1}$ to DAC to obtain the next $R_{ADC}$ bits of $x$.
After $\left\lceil Q_x/R_{ADC}\right\rceil$ iterations, we can obtain all the $Q_x$ bits of vector $x$.
In every loop, the matrix $A$ participates in a VMM computation.
This VMM is also carried out by the INV crossbars storing $A$.
We only need to connect these INV crossbars to the DAC/ADC interface and the crossbars are able to calculate VMM.

\begin{figure*}
  \centering
  \includegraphics[width=0.95\linewidth]{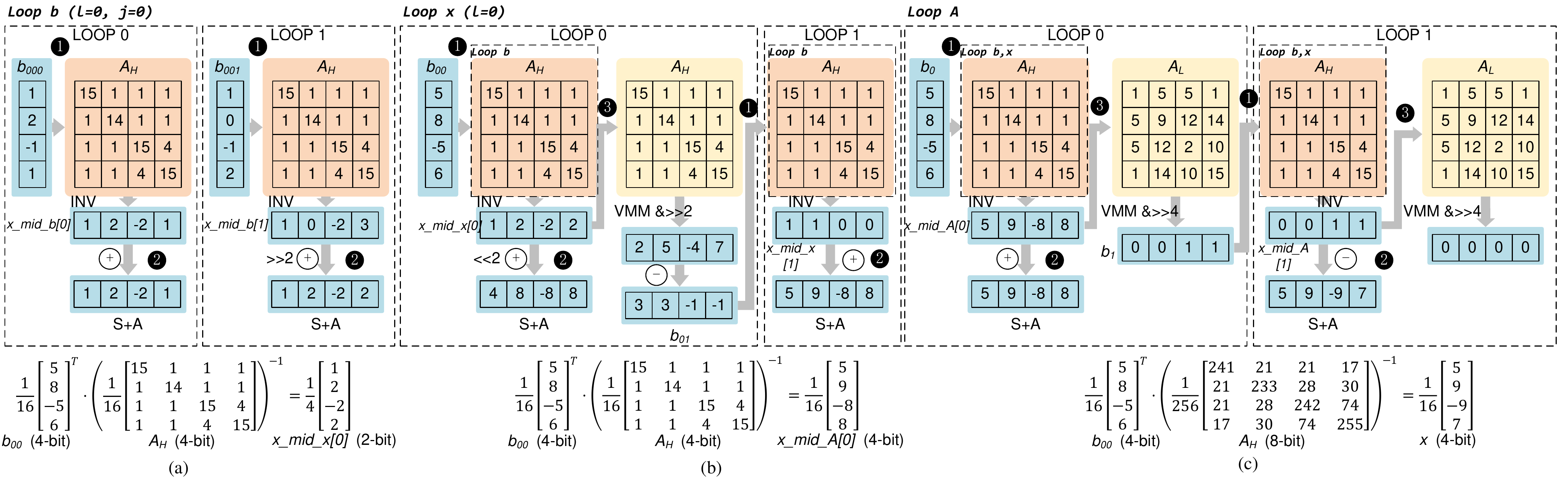}
\vspace{-10pt}
\caption{An example of high-precision matrix inversion algorithm on ReRAM crossbar.
  (a) \textit{Loop b} when $l=j=0$.
  (b) \textit{loop x} when $l=0$.
  (c) \textit{loop A}.
  Matrix $A$ is 8-bit, input $b$ and result $x$ are 4-bit. DAC and ADC resolution is 2-bit. ReRAM cell can write 4-bit.}
  \label{fig:mat_inv}
\vspace{-15pt}

\end{figure*}

\textit{3) Taylor Expansion for matrix $A$.}
\revisionadd{
  For matrix $A$, we cannot directly split it into multiple slices according to $R_c$.
This is because there is no distributive law on $A$: $x=\left(A_1+A_2\right)^{-1}\cdot b\neq A_1^{-1}\cdot b+A_2^{-1}\cdot b$.}
Therefore, we choose another route to address the problem and propose a method based on Taylor Expansion.
We split the matrix $A$ into two slices: the first $k\cdot R_c$ bits $A_H$, and the rest $Q_A-k\cdot R_c$ bits $A_L=\left(A-A_H\right)\cdot 2^{kR_c}$.
$A_H$ is written into the INV crossbars and $A_L$ is written into a separate VMM crossbar as shown in \figurename{}~\ref{fig:mat_inv}.
In the second-order training algorithms, $A$ is a symmetric matrix \cite{quasinewtonnocedal2006numerical}.
Therefore, both $A_H$ and $A_L$ are also symmetric metrics and the matrix multiplication is commutative.
We can simplify the differential of matrix $A_H$ as:
\begin{eqnarray}
    \mathrm{d}\left(A_H^{-1} \right) & = & -A_H^{-1}\cdot \mathrm{d}A_H\cdot A_H^{-1}=-A_H^{-2}\cdot \mathrm{d}A_H
\end{eqnarray}
According to Taylor Expansion, we have:
\begin{eqnarray}
A^{-1}\cdot b & = & \left(A_H+A_L\cdot2^{-kR_c}\right)^{-1}\cdot b\\
    & = & A_H^{-1}\cdot\left(I - P + P^2-\cdots\right)\cdot b
    \label{eqn:taylor}
\end{eqnarray}
where $P=A_H^{-1}\cdot A_L\cdot2^{-kR_c}$.
In theory, we can obtain an accurate final result by computing all the terms on the right-hand side of Eqn. \ref{eqn:taylor}.
In reality, we only need to compute a small number of terms in order to get the desired precision.
Suppose the $l^{\mathrm{th}}$ term is $b_l=A_H^{-1}(-P)^{l}b$.
Note that adding one more term only requires one more multiplication with $P$, i.e. $b_{l+1}=-P\cdot b_l$, which is essential to loop on the INV crossbars one more time (for $\cdot A_H^{-1}$), and one more time on the VMM crossbars (for $\cdot A_L$).

\begin{figure}
  \centering
\includegraphics[width=0.9\linewidth]{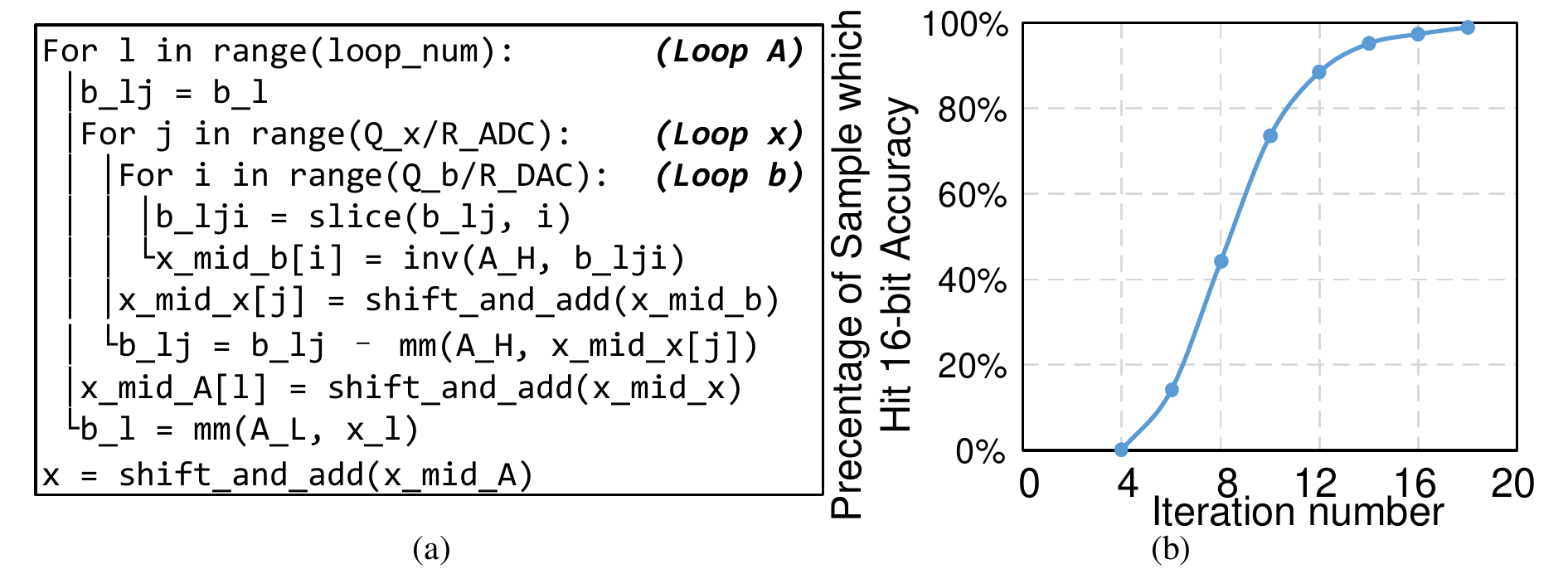}
\vspace{-10pt}
\caption{(a) Pseudo-code of high-precision matrix inversion algorithm based on low-precision matrix inversion operation. (b) The number of Taylor expansion iteration required to achieve 16-bit accuracy in matrix inversion. The matrix's size is $1024\times1024$. Matrix, input vector and result are all 16-bit quantized.}
  \label{fig:pseudo}
  \vspace{-14pt}
\end{figure}

Pseudo-code of combining three steps into one complete high-precision matrix inversion is summarized in \figurename{}~\ref{fig:pseudo} (a).
The three nested loops, \textit{Loop b}, \textit{Loop x} and \textit{Loop A} correspond to the bit split of $b$, $x$ and $A$.
To better demonstrate the implementation of the three schemes described with Eqn.~\ref{eqn:b-slice} and \ref{eqn:taylor}, we give an example in \figurename{}~\ref{fig:mat_inv}.
Input vector $b$ and result vector $x$ are 4-bit quantized.
The matrix $A$ is 8-bit quantized and split into two 4-bit slices $A_H$, $A_L$.
$A_H$ is programmed to an INV crossbar and $A_L$ is written to a VMM crossbar.
Both ADC and DAC are 2-bit, therefore, both \textit{Loop b} and \textit{Loop x} have two iterations.
\figurename{}~\ref{fig:mat_inv}(a) shows \textit{Loop b} when $l=0, j=0$.
The input 4-bit vector $b_{0,0}$ is splited into two 2-bit slices $b_{0,0,0}$ and $b_{0,0,1}$ according to Eqn. \ref{eqn:b-slice}.
The two slices are successively input to the INV crossbar and the results are accumulated with S+A circuit.
\figurename{}~\ref{fig:mat_inv}(b) shows \textit{Loop x} when $l=0$.
In loop 0, we first compute the matrix inversion with 4-bit vector $b_{0,0}=b_0$ and $A_H$, and obtain 2-bit result $x\_mid\_x[0]$.
After that, we input $x\_mid\_x[0]$ to the VMM crossbar and compute with $A_H$.
Result of the VMM is subtracted from $b_{0,0}$ and obtain the input of the next loop $b_{0,1}$.
The results of two iterations $x\_mid\_x[0,1]$ are accumulated with S+A circuit to obtain a 4-bit result.
\figurename{}~\ref{fig:mat_inv}(c) shows the ouside \textit{Loop A}.
In the first loop, we first compute the matrix inversion between the origin 4-bit vector $b_0=b$ and $A_H$, and obtain the result $x\_mid\_A[0]$.
$x\_mid\_A[0]$ is correspond to the first term in Eqn. (9) ($A_H^{-1}b$), therefore, it is added with the accumulated result by S+A.
Then, $x\_mid\_A[0]$ is input to $A_L$ and obtain the input of next loop $b_1$.
In the second loop, the intermediate result $x\_mid\_A[1]$ corresponds to the second term of Eqn.(9) ($A_H^{-1}Pb$).
Therefore, it is subtracted from the accumulated result.
After two loops, the result of VMM $b_2$ is a zero vector.
This means that the accumulated result will not change in the following loops and the accurate 4-bit result is achieved.

Then we present the computing cycle spent on one high-precision matrix inversion.
\textit{Loop b} has $\left\lceil Q_b/R_{DAC}\right\rceil$ matrix inversion computations.
\textit{Loop x} has $\left\lceil Q_x/R_{ADC}\right\rceil$ loops and each loop contains one \textit{Loop b} and one VMM computation.
Each loop of \textit{Loop A} contains one \textit{Loop x} and one VMM computation.
Suppose \textit{Loop A} has $N$ loops,
the total cycles spent on one high-precision matrix inversion is:
\begin{eqnarray}
c_{INV} & = & N \left(2\left\lceil\dfrac{Q_b}{R_{DAC}} \right\rceil\cdot\left\lceil\dfrac{Q_x}{R_{ADC}}\right\rceil + \left\lceil\dfrac{Q_x}{R_{DAC}}\right\rceil\right)
\end{eqnarray}

Unlike the bit-slicing scheme for $x$ and $b$, the iteration number for computing $A$ is not decided by hardware parameters.
\revisionadd{
This is because the convergence of Taylor Expansion is related to the matrix $A$ itself.
The Taylor Expansion method is efficient only if the following two conditions are satisfied:
First, $A_H^{-1}$ is closed to $A^{-1}$.
Second, the Taylor Equation can converge fast.
It can be proved that both conditions have a common sufficient condition, which is that $A$'s condition number $\kappa \left(A\right)$ need to be small \cite{meyer2000matrix}. 
}
Fortunately, in the second-order optimization algorithms, Tikhonov regularization has been introduced to the SOI matrices \cite{hessianfreekiros2013training,distributed8850515,kfacpmlr-v37-martens15}.
This method can largely reduce the condition number of the matrix $A$ and makes the Taylor Equation converge faster.


\subsection{Verification of the Proposed High-Precision Matrix Inversion}


To confirm the correctness of our proposed high-precision matrix-inversion, we write a RTL model with Verilog HDL for the circuit proposed in \figurename{}~\ref{fig:vmm_inv}.
As the ReRAM crossbar can not be directly modeled with Verilog, we write behavioral models for the low-precision INV and VMM crossbars with 4-bit precision and integrate them into the Verilog code.
The non-idealism of OpAmp, including limited input resistance, output resistance and gain, is compensated with the method proposed in \cite{predictioner9407108} and properly modeled.

We generate \revisionadd{$10^6$} input test vectors to verify the correctness of the proposed high-precision matrix inversion circuit in Verilog.
The size of the input matrices is $1024\times1024$.
Tikhonov Normalization of the same level of ResNet 50 training is applied to the matrices.
All the samples can achieve the required 16-bit accurate result after enough iterations.
The relation between the percentage of samples that can achieve 16-bit accuracy and the loop number of \textit{Loop A} is plotted in \figurename{}~\ref{fig:pseudo}(b).
Over 99\% samples can achieve 16-bit accuracy within 18 loops.
Therefore, in the following sections of the paper, we set the loop number of \textit{Loop A} in \figurename{}~\ref{fig:pseudo} (b) to 18.

\vspace{0.1cm}
\section{Architecture}
\label{sec:architecture}


\begin{figure}
  \centering
  \includegraphics[width=0.8\linewidth]{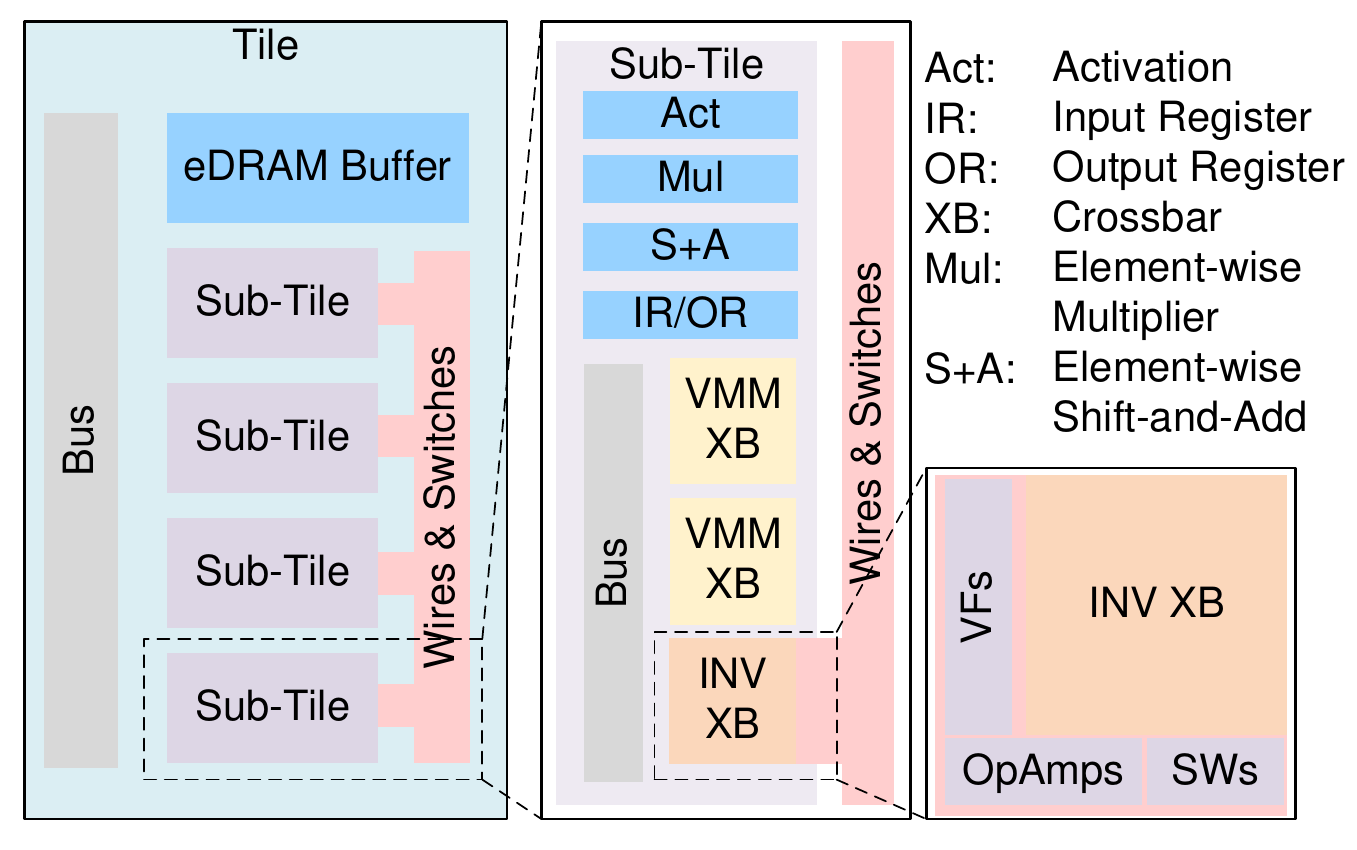}
\vspace{-10pt}
\caption{Architecture overview of \archname{}.}
  \label{fig:arch_overview}
  \vspace{-20pt}
\end{figure}

Based on the high-precision matrix inversion presented in Section \ref{sec:matinv}, we propose \archname{}, a ReRAM-based accelerator architecture for the second-order training.
\archname{} is composed of multiple tiles connected through an on-chip network.
The structure of each tile is shown in \figurename{}~\ref{fig:arch_overview}.
A tile contains a buffer, a bus for data movement, and several sub-tiles.
In each sub-tile, there are one INV crossbar and several VMM crossbars for calculating matrix multiplication and inversion, as well as some other auxiliary circuits including input/output registers (IR/OR), activation function (Act), S+A and multipliers (Mul).
The element-wise operations, including batch normalization and activation function, are processed by Acts, S+A and Muls.
The ratio of VMM crossbars to INV crossbar in a sub-tile will be explained in Section \ref{evaluation:result:dse}.
All the INV crossbars in one tile are connected by wires \& switches.
The connection is described in Section~\ref{sec:architecture:invconnection}. The data communication between INV and VMM crossbars is through internal registers or eDRAM buffers.

We have summarized that there are two major operators in the second-order training algorithms: matrix multipliaction (MM) and INV.
The VMM crossbars in a sub-tile will be responsible for the computation of MM.
The INV operation is jointly handled by VMM and INV crossbars through multiple computing loops as the scheme proposed in Section~\ref{sec:matinv}.

\subsection{Connecting Multiple INV Crossbars}
\label{sec:architecture:invconnection}

\begin{figure*}
\includegraphics[width=0.95\linewidth]{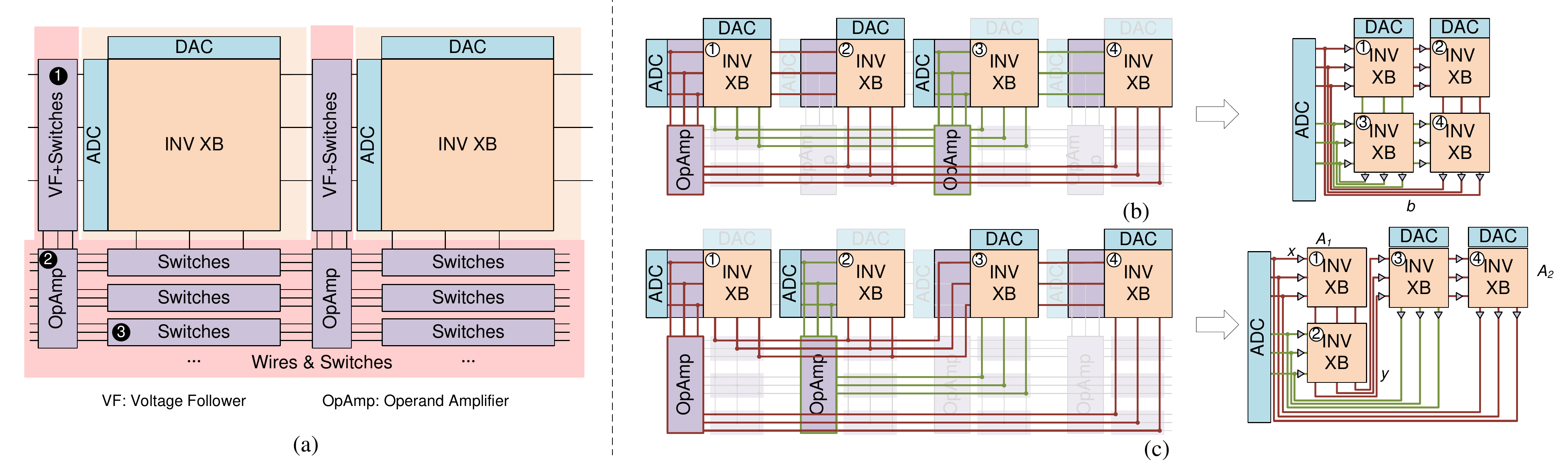}
\vspace{-10pt}
  \caption{The ReRAM crossbars for INV.
    (a) Multiple crossbars are in one sub-tile and can be configured.
    (b) The connection configuration when computing a $2\times2$ larger INV, and the equivalent circuit.
    (c) The connection configuration when fusing a VMM and an INV, and the equivalent circuit.}
  \label{fig:inv_element}
\vspace{-10pt}
\end{figure*}

In the second-order training, the SOI matrix is usually larger than the size of a single INV crossbar.
For example, one of VGG-19 layers has $512$ input channels and the kernel size is $3\times3$.
When training this network with K-FAC algorithm, the SOI matrix $A$ of this layer is $4608\times4608$.
But the typical size of a ReRAM crossbar is $256\times256$, which is smaller than the SOI matrix.
The VMM operation in high-precision matrix inversion can be performed by splitting the matrix to multiple VMM crossbars and adding up the mid-summations with digital adders \cite{prime10.1109/ISCA.2016.13}.
While for the INV operation, the INV crossbars can only compute as one analog circuit.
Therefore, we need to connect multiple INV crossbars so as to form a combined structure for large matrix inversion computation.
Moreover, as the size of SOI matrix can be different over different networks, reconfigurability is required to change the connections.

The connection scheme for INV crossbars in a tile is depicted in \figurename{}~\ref{fig:inv_element}(a).
Around each INV crossbar, there is a group of VFs \blackcir{1}, a group of OpAmps \blackcir{2} and a group of switches (SWs) \blackcir{3}.
The VFs are connected to the wordlines of the INV crossbar and can be configured to receive the analog signals from the left crossbar or from the OpAmps as inputs.
SWs are connected to the bitlines of the INV crossbar.
SWs and OpAmps are connected by a group of wires.
If SWs and OpAmps in the same INV crossbar are configured to connect, the analog signals from the crossbar's bitlines can be applied to the OpAmps as inputs.

We use an example to explain how to configure the connections for computing a large-size matrix inversion.
Suppose matrix $A$'s size is $512\times512$ and the INV crossbar size is only $256\times256$.
This means that the matrix needs to occupy $2\times2$ INV crossbars.
The configuration is shown in \figurename~\ref{fig:inv_element}(b).
The VFs of crossbar \whitecir{2} and \whitecir{4} are configured to accept the signals from crossbar \whitecir{1} and \whitecir{3}, respectively.
This makes crossbar \whitecir{1} and \whitecir{2} (or \whitecir{3} and \whitecir{4}) share the same analog voltage input.
The bitlines of crossbar \whitecir{1} and \whitecir{3} are connected (shorted) by configuring their SWs and fed to the inputs of OpAmps in crossbar \whitecir{3}.
This means that the current from bitlines in crossbar \whitecir{1} and \whitecir{3} are accumulated and fed back to the wordlines of crossbar \whitecir{3}.
Similarly, the bitlines of crossbar \whitecir{2} and \whitecir{4}, and the inputs of OpAmps in crossbar \whitecir{1} are connected, effectively forming an enlarged INV crossbar circuit structure.
The input vector $b$ can be applied to the DACs of INV crossbar \whitecir{1} and \whitecir{2}, and the results $x$ are generated in the ADCs of crossbar \whitecir{1} and \whitecir{3} as shown in the equivalent circuitry in \figurename{}~\ref{fig:inv_element}(b).
Therefore, the four crossbars are combined to function as one $512\times512$ matrix inversion circuit.


The maximum size of matrix inversion is limited by the total number of INV crossbars in a tile, which is an important architectural design parameter.
Nevertheless, most second-order training algorithms allow us to flexibly tune the SOI matrix size to achieve different trade-offs between accuracy and cost.
So we can always use the proper SOI matrix sizes to fulfill the limitation of INV crossbars, meanwhile keeping convergence rate in check.

\subsection{Fused Operation}
\label{sec:architecture:fuse}

The circuit organization described in \figurename{}~\ref{fig:inv_element}(a) makes it capable of computing a fused operation of low-precision matrix inversion and its immediately leading low-precision matrix multiplication.
Suppose we need to calculate a fused-matrix-multiplication-and-inversion $x=(A_1\cdot A_2)^{-1}b$, in which $A_1$ has a size of $512\times256$ and $A_2$ has a size of $256\times512$.
Therefore, both $A_1$ and $A_2$ occupy two crossbars.
The connection is depicted in \figurename{}~\ref{fig:inv_element}(c).
The four INV crossbars are divided into 2 groups.
Matrix $A_1$ is mapped to crossbar \whitecir{1} and \whitecir{2}, and matrix $A_2$ is mapped to crossbar \whitecir{3} and \whitecir{4}.
The bitlines of crossbar \whitecir{1} and \whitecir{2} are connected so that their currents are accumulated.
The accumulated currents are applied to the wordlines of crossbar \whitecir{3}.
The VFs of crossbar \whitecir{4} is configured to share the same inputs with crossbar \whitecir{3}.
The bitlines of crossbar \whitecir{3} is connected to the inputs of OpAmps to feedback to the wordlines of crossbar \whitecir{2}.
Similarly, the signals on the bitlines of crossbar \whitecir{4} are fed back to the wordlines of crossbar \whitecir{1}.
The equivalent circuitry is shown in \figurename{}~\ref{fig:inv_element}(c).
The analog signals (representing input $x$) on the wordlines of crossbar \whitecir{1} and \whitecir{2} are multiplied with the matrix $A_1$, and the results on bitlines can be represented by $y=A_1\cdot x$.
The signal $y$ is applied to the wordlines of crossbar \whitecir{3} and \whitecir{4}, and the results are fed back to the input $x$.
When we apply input $b$ to the DACs in the crossbar \whitecir{3} and \whitecir{4}, the relation between $b$ and $y$ is given by $b=A_2\cdot y$.
Since $y=A_1\cdot x$, therefore, $b=A_1\cdot A_2\cdot x$, and in turn $x=(A_1\cdot A_2)^{-1}b$.
In this way, the circuit can carry out a low-precision fused-matrix-multiplication-and-inversion.

The high-precision matrix inversion scheme can also be applied to the fused operation.
Suppose the matrix $A_1$ and $A_2$ are both 16-bit quantized. The higher 8 bits of $A_1$ is $A_{1H}$ and $A_{1L}=\left(A_1-A_{1H}\right)\cdot2^8$.
Similarly, the higher 8 bits of $A_2$ is $A_{2H}$ and $A_{2L}=\left(A_2-A_{2H}\right)\cdot2^8$.
Denote $A_H=A_{1H}\cdot A_{2H}$ and we have:
\begin{eqnarray}
A_L & = & \left(A-A_H\right)\cdot2^8=\left(A_{1}A_2-A_{1H}A_{2H}\right)\cdot2^8\\
  &=&A_{1H}A_{2L}+A_{1L}A_{2H}+A_{1L}A_{2L}\\&=&A_{1}\cdot A_{2L}+A_{1L}\cdot A_{2H}
  \label{eqn:AL}
\end{eqnarray}
According to Eqn. \ref{eqn:taylor}, $A_H$ only participates in INV and $A_L$ only participates in VMM.
Therefore, $A_{1H}$ and $A_{2H}$ are programmed to the INV crossbars. $A_{1}$, $A_{1L}$, $A_{2H}$ and $A_{2L}$ are programmed to the VMM crossbars.
The matrix multiplication between input vector $b$ and matrix $A_L$ is computed via multiple VMMs according to Eqn. \ref{eqn:AL}.
Note that the two VMM terms can be computed in parallel, and each term contains two VMM operations (e.g. $A_{1}\cdot A_{2L}\cdot b$).
The total cycle is:
\begin{eqnarray}
c_{INV+VMM}  =  N \left(2\left\lceil\dfrac{Q_b}{R_{DAC}} \right\rceil\cdot\left\lceil\dfrac{Q_x}{R_{ADC}}\right\rceil + 2\left\lceil\dfrac{Q_x}{R_{DAC}}\right\rceil\right)
\end{eqnarray}

\vspace{0.1cm}
\section{Mapping scheme}
\label{sec:mapping}
\subsection{Dataflow Graph (DFG) Partition}
\label{sec:mapping:dfg}

\begin{figure}
    \includegraphics[width=0.95\linewidth]{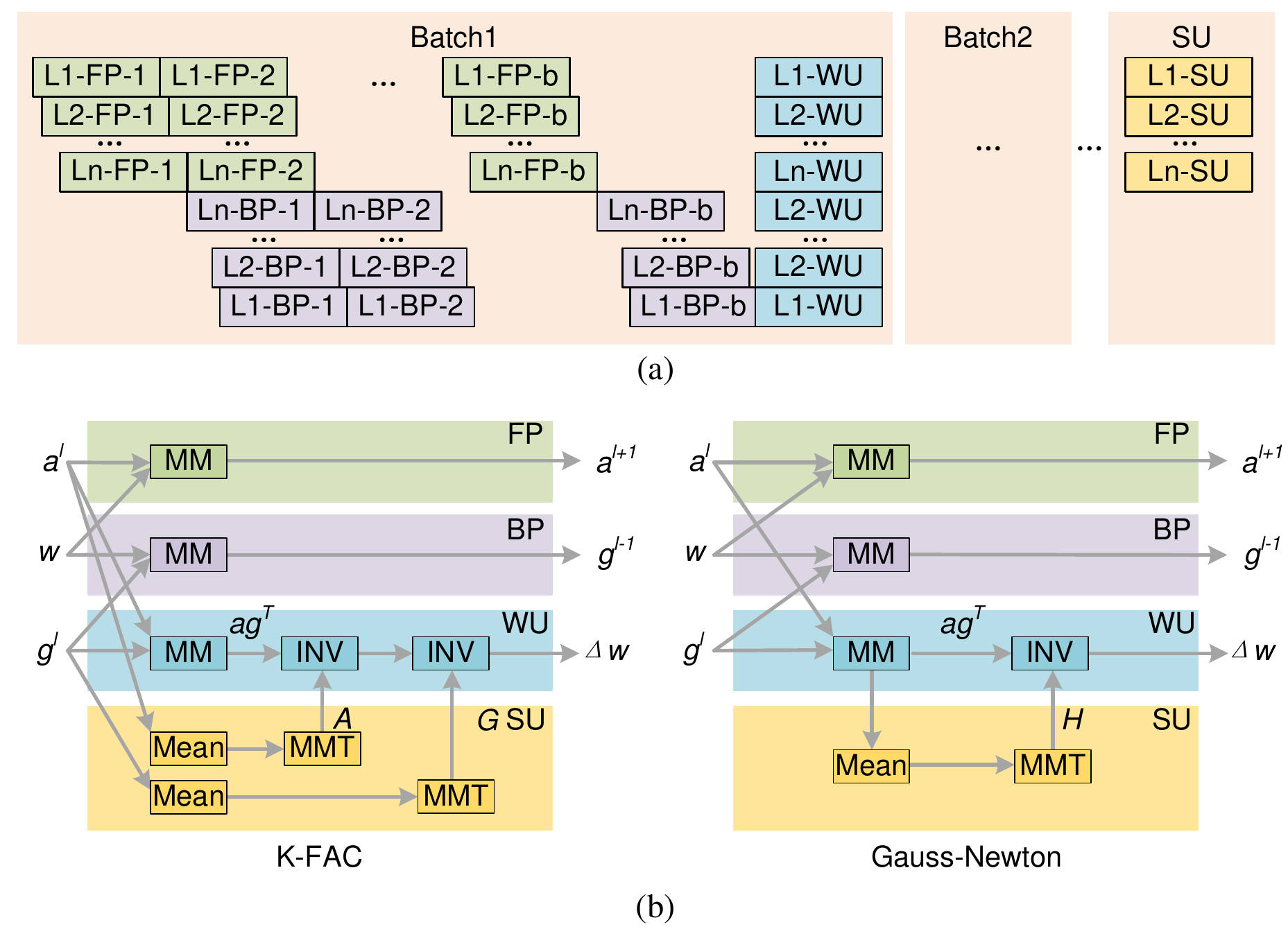}
\vspace{-7pt}
\caption{(a) The pipeline of \archname{}. `L2-FP-1' represents the second layer's FP graph is executed on the first input.  (b) The graph partition of two second-order training algorithms, K-FAC and Gauss-Newton. MM: Matrix Multiplication, MMT: Matrix Multiply with its Transpose}
    \label{fig:pipeline}
    \vspace{-15pt}
\end{figure}

In the second-order training algorithms, the calculation of each DNN layer can be partitioned into four DFGs:
Forward Propagation (FP), Backward Propagation (BP), Weight Update (WU), and SOI Update (SU).
As shown in \figurename~\ref{fig:pipeline}, for each batch, the FP graph is firstly executed on each input and obtain the error as output.
During FP, the layers are calculated with inter-layer pipelining \cite{ISAAC} to reduce latency.
Whenever the FP on one input has been done, the BP graph starts.
The WU graph only executes once after an entire batch has been processed.
$\Delta w$ values are computed and the weight matrices are updated.
For every a few batches, the SU graph executes once to update the SOI matrices.

Take the two state-of-the-art second-order optimization methods, KFAC \cite{kfacpmlr-v37-martens15} and Gauss-Newton \cite{hessianfreekiros2013training} as examples.
In the FP graph, the input feature map $a$ is multiplied with weight $w$ to obtain the output feature map.
In the BP graph, the error $g$ is multiplied with weight $w$ and back-propagated to the previous layer.
In the WU graph, both of the two algorithms firstly compute the first-order gradient of weights.
Then the gradients of weights multiply with the SOI matrix $A$, $G$, or $H$ to obtain the  $\Delta w$.
In the SU graph, the SOI matrices are updated with the rule described in Section \ref{sec:background:so_algorithms}.

\subsection{Pattern Mapping Scheme}
\label{sec:mapping:pattern}

\begin{figure*}
    \centering
\includegraphics[width=\linewidth]{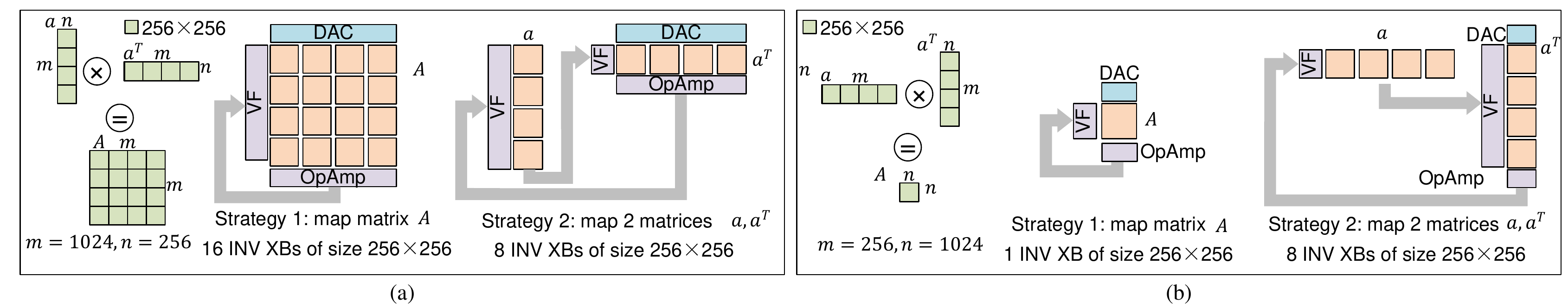}
\vspace{-10pt}
    \caption{Two cases when mapping the SU graph. The inversion of matrix $A$ is needed and $A$ is obtained by $A=a\cdot a^T$ (a) When $m\gg n$, strategy 2 can achieve a smaller crossbar occupation. (b) When $m\ll n$, strategy 1 is better.
    }
    \label{fig:mapping}
\end{figure*}

We propose a mapping scheme for the two typical patterns in the SU graph to improve the performance of \archname{}.
The first is the MM-INV pattern, where the matrix for inversion is obtained after several matrix multiplication (MM) operators.
The second pattern is composed of several successive MM or INV operators.

\subsubsection{MM-INV pattern}
\label{sec:mapping:pattern:mm-inv}

This pattern can be frequently found in the SU graph of the two algorithms.
The VMM operation is followed by a matrix inversion.
Denote the matrix for inversion as $A\in \mathbb{R}^{m\times m}$, which is the product of matrix $a \in \mathbb{R}^{m\times n}$ and its transpose, $A = a\cdot a^T$.
The matrix $A$'s inversion is multiplied with vector $b$ to obtain the final result $x = (a\cdot a^T)^{-1}\cdot b$.

We can use two different strategies to map this pattern onto the crossbars, and they exhibit different crossbar occupations and computational costs with different matrix sizes.
In the first strategy, $A=a\cdot a^T$ is calculated first using VMM crossbars and then mapped to INV crossbars to compute $x$.
In the second strategy, $A$ is not explicitly computed; instead, $a$ and $a^T$ are written to the INV crossbars to perform a fused-matrix-multiplication-and-inversion as described in Section IV-B.

Neither of the two strategies can always achieve the lowest crossbar occupation.
In the first case, as shown in \figurename{}~\ref{fig:mapping}(a), where the size of matrix $a$ is $1024\times 256$.
The matrix $A$ is the product of $a\cdot a^T$ so that its size will become $1024\times1024$.
Suppose the size of an INV crossbar is $256\times256$.
In the first strategy, $A$ is mapped directly to the INV crossbars and occupies $16$ INV crossbars.
In the second strategy, either $a$ or $a^T$ occupies $4$ INV crossbars, taking up a total of $8$ INV crossbars.
In comparison, the second strategy takes only half of the crossbars compared to the first one.
As in Eqn. 14, the second strategy have a longer latency.
However, the latency grows less than $2\times$ compared to the first strategy but the overall performance is still better due to the much-reduced resource consumption.

In another case, as illustrated in \figurename{}~\ref{fig:mapping}(b), where the size of matrix $a$ is $256\times 1024$ so that the
size of matrix $A$ will be $256\times256$ and occupy only $1$ INV crossbar. Therefore the first strategy should be more efficient since the second strategy will still need 8 crossbars.

With the observation above, we propose a smart mapping scheme for the MM-INV pattern.
We can first locate an INV node in the DFG and then search backward.
If the predecessor node is an MM, we need to decide whether to fuse the node with INV or not, by comparing a cost function of the two mapping strategies.
Suppose the two matrices for MM is $m\times n$ and $n\times k$, and the size of an INV crossbar is $s\times s$.
If we choose to fuse the nodes (the second strategy), the cost function is:
\begin{eqnarray}
C_{fuse} = \alpha\cdot c_{VMM+INV}+\beta\left(\left\lceil\dfrac{n}{s}\right\rceil\cdot\left(\left\lceil\dfrac{m}{s}\right\rceil+\left\lceil\dfrac{k}{s}\right\rceil\right)\right)
\end{eqnarray}
The first term represents the computational latency for the fused-matrix-multiplication-and-inversion, and the second term describes the condition of crossbar occupation.
$\alpha$ and $\beta$ are two empirical coefficients to trade off performance and resource consumption.
Similarly, if we do not fuse the nodes (the first strategy), the cost function becomes:
\begin{eqnarray}
C_{non-fuse} = \alpha\cdot c_{INV}+\beta\left(\left\lceil\dfrac{m}{s}\right\rceil\cdot\left\lceil\dfrac{k}{s}\right\rceil\right)
\end{eqnarray}
By simply comparing the two cost functions, we can decide which strategy should be adopted.


\subsubsection{Successive MM/INV pattern}
\label{sec:mapping:performance:latency}

Different from the SU graph, the WU phase exhibits a different pattern.
Take the WU graph of the K-FAC algorithm as an example.
The weight update is calculated as $\Delta w=A^{-1}\cdot(a\cdot g^T)\cdot G^{-1}$.
Consider a convolution layer with input channel, output channel, input feature map size and kernel size as $c_{in}$, $c_{out}$, $h\times w$ and $k\times k$.
$A\in\mathbb{R}^{c_{in}k^2\times c_{in}k^2}$, $G\in\mathbb{R}^{c_{out}\times c_{out}}$, $a\in\mathbb{R}^{c_{in}k^2\times hw}$ and $g\in\mathbb{R}^{c_{out}\times hw}$.
We also have two strategies to compute the formula.
In the first strategy, we firstly compute $p=a\cdot g^T$ using VMM crossbars.
As input feature map $a$ is generated in FP phase and afterwards, the $g$ is computed in BP phase.
When waiting for $g$, we can program $a$ to VMM crossbars.
If $a$ is larger than the size of a VMM crossbar, $a$ is mapped to multiple crossbars with the splitting scheme in \cite{prime10.1109/ISCA.2016.13} and these crossbars can be programmed in parallel.
Therefore, the crossbar programming latency can be completely covered by the FP and BP phase.
After $g$ is generated, we send $g$ row by row to the crossbars, and this computation consumes $c_{out}\cdot c_{VMM}$ cycles.
Then we compute $q=A^{-1}\cdot p$ by applying $p$ row by row to the INV crossbars that store $A$.
As $p\in \mathbb{R}^{c_{in}k^2\times c_{out}}$, this procedure needs $c_{out}\cdot c_{INV}$ cycles.
Finally, $\Delta w=q\cdot G^{-1}$ is computed by applying $q$ row by row to the INV crossbars that store $G$, which takes $c_{in}k^2\cdot c_{INV}$ cycles.
As the first step and the second step can be performed in a pipeline, this strategy totally consumes $\left(c_{in}k^2 + c_{out}\right)\cdot c_{INV}+c_{VMM}$ cycles.

In the second strategy, we firstly compute $r=A^{-1}\cdot a$ by input $a$ to the INV crossbars that store $A$, and write $r$ to VMM crossbars.
This computation can be executed immediately when $a$ is generated in FP phase, and the latency can be covered because it will take a while for $g$ to come back in the BP phase.
Then we compute $s=g^T\cdot G^{-1}$ by input $g$ to the INV crossbars that store $G$, which needs $hw\cdot c_{INV}$ cycles.
Finally, $\Delta w=r\cdot s$ is computed by input $s$ to the VMM crossbars that store $r$, and takes $c_{out}\cdot c_{VMM}$ cycles.
This strategy needs $hw\cdot c_{INV}+c_{out}\cdot c_{VMM}$ cycles in total.

The two strategies are only different in computing latency, therefore, the cost functions of the two strategies are defined as the computing latency.
For each layer, we compute the cost function of the two strategies and choose a faster strategy.
In typical CNNs, for the first a few layers, the feature map size is usually large while the number of input or output channels is small.
Therefore, the first strategy may spend fewer cycles to achieve a better performance.
For the other layers, the feature map size becomes smaller with an increasing number of input and output channels, and the second strategy should be the better choice with reduced execution latency.


\vspace{0.1cm}
\section{Evaluation}
\label{sec:evaluation}
\subsection{Experiment setup}
\label{evaluation:setup}

We evaluate our proposed \archname{} with seven large-scale CNNs (VGG-13, VGG-16 and VGG-19 \cite{vggsimonyan2015deep}, MSRA-1 and MSRA-2 \cite{msra7410480}, ResNet-50 and ResNet-101 \cite{resnethe2015deep}) on ImageNet dataset \cite{imagenet_cvpr09}, BERT \cite{devlin2019bert} on MLPerf dataset \cite{mattson2019mlperf}, and autoencoder \cite{autoencoderdoi:10.1126/science.1127647} on MINST dataset\cite{lecun2010mnist}.
The batch size is set to 256, and the SOI matrices are updated after every 10 batches.
We present the minimum and maximum SOI matrix sizes in \tablename{}~\ref{tab:DNN_param}.

\begin{table}
    \centering\small
\caption{Min/Max SOI matrix size of DNN Benchmarks. Convolution layers are formatted as C$k$x$k$,$c_{in}$/$c_{out}$. SOI matrix size $b$B+$r$ means $b$ blocks of size $1024\times 1024$ and one block of size $r\times r$}
\vspace{-5pt}
\label{tab:DNN_param}

\resizebox{0.8\linewidth}{!}{
    \begin{tabular}{|c|c|c|c|c|}
        \hline
        Network                      &     & Layer          & SOI Matrix           \\ \hline
\multirow{2}*{\begin{tabular}{c}VGG- \\ \revisionadd{13/}16/19\end{tabular}}     & Min & C3x3, 3/64     & A: 0B+27, G: 0B+64   \\ \cline{2-4}
                                     & Max & C3x3, 512/512  & A: 4B+512, G: 0B+512 \\ \hline
\multirow{2}*{\begin{tabular}{c}MSRA-\\\revisionadd{1/}2\end{tabular}}        & Min & C7x7, 3/96     & A: 0B+147, G: 0B+96  \\ \cline{2-4}
                                     & Max & C3x3, 512/512  & A: 4B+512, G: 0B+512 \\ \hline
\multirow{2}*{\begin{tabular}{c}ResNet-\\50/101\end{tabular}} & Min & C1x1, 64/64    & A:0B+64, G: 0B+64    \\ \cline{2-4}
                                     & Max & C3x3, 512/2048 & A: 4B+512, G:0B+512  \\ \hline
\multirow{2}*{Bert} & Min & Projection    & A:0B+768, G: 0B+64    \\ \cline{2-4}
& Max & Feed Forward & A: 3B+0, G:0B+768  \\ \hline
\end{tabular}}
\end{table}

We compare the \archname{} architecture with NVidia Tesla V100 GPU and one recent PIM architecture for the first-order training, PipeLayer \cite{pipelayer7920854}.
Second-order training algorithm K-FAC \cite{kfacpmlr-v37-martens15} is executed on GPU and \archname{} and SGD is executed on GPU and Pipelayer.
To evaluate the energy consumption and latency of GPU, we train the benchmarks with Mindspore framework \cite{mindspore}.
We build a cycle-accurate simulator to evaluate \archname{}'s area and energy consumption.
We adopt the ReRAM crossbar energy/area model in \cite{dot7544263}, ADC energy/area model in \cite{adc6607254}, and DAC energy/area model in \cite{dac5711005}.
The resolution of ADC and DAC is 8-bit and 4-bit respectively.
The OpAmp energy/area model is from \cite{OpAmp}.
The eDRAM buffer and bus are simulated with CACTI7.0 \cite{cacti10.1145/3085572}.
Hyper Transport link for inter-chip communication use the model from \cite{ISAAC}.
All the circuits are scaled to 28nm technology.
\revisionadd{To make a fair comparation with PipeLayer, we use the same model for the common circuits of the two architectures, such as ReRAM crossbars. The RePAST tiles’ computation and memory access are controlled by a state machine. We write a simple optimizer to generate the state machine according to the DNN’s hyper-parameters.}
Energy consumption of GPU is evaluated with $\mathtt{nvidia}$-$\mathtt{smi}$ tool.
The ReRAM crossbar size in \archname{} is set to $256\times 256$ and each ReRAM cell can support 4-bit precision. We always assign one INV crossbar in a sub-tile.
In order to support a wider range of block size in the second-order training algorithms, we set the total number of INV crossbars (or sub-tiles) in a tile as 16 so that a maximum block size of $1024 \times 1024$ can be supported.
This block size is larger than most layer's SOI matrices and these matrices are not approximated in the algorithm.
For example, over 80\% of ResNet-50's SOI matrix is smaller than this size.
Hyper-parameters of the mapping scheme are set to $\alpha=1$ and $\beta=0.1$.

\revisionadd{
    \textbf{Cycle Time.}
    We set the cycle time of \archname{}'s ReRAM crossbars as 100 ns.
    This value is commonly used in ReRAM-based accelerators \cite{ISAAC,Timely,NeuralPIM} because it is coherence to current ADC frequency.
    For INV crossbars, the settling time is proportion to the $\lambda_{min}^{-1}$ of matrix, and OpAMP parameters $L_0^{-1}\omega_0^{-1}$, and is independent to the size of Matrix size.
    \cite{predictioner9407108}'s INV crossbar can settle down within 50 ns.
    We use the same OpAMP as \cite{predictioner9407108}, and the Tikhonov normalization in second-order training can reduce the $\lambda_{min}^{-1}$ value of matrix.
    Therefore, our INV corssbar can settle down within 100 ns.
}

\subsection{Design Space Exploration}
\label{evaluation:result:dse}
\begin{figure}
    \centering
\includegraphics[width=0.7\linewidth]{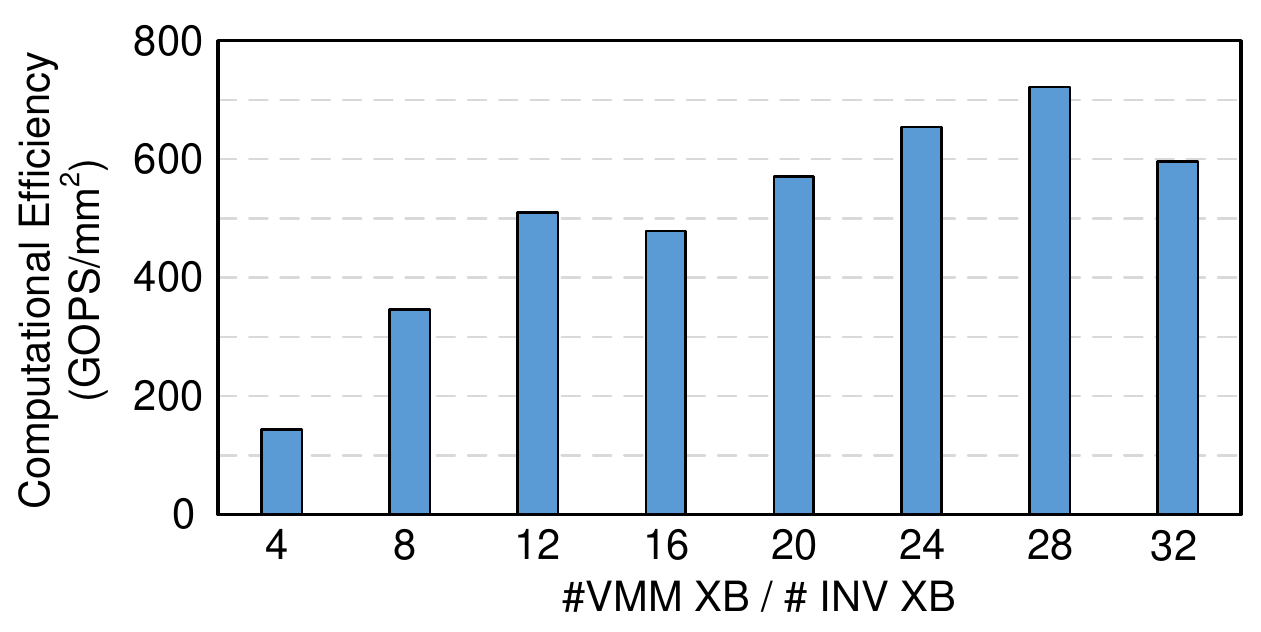}
\vspace{-10pt}
    \caption{ Design space exploration.  We search for the best proportion between VMM and INV crossbar number with average computational efficiency across benchmarks as metric.}
    \label{fig:result_dse}
\vspace{-10pt}
\end{figure}

One important design parameter of the RePAST architecture is the proportion between the INV crossbar number and the VMM crossbar number in a tile.
On the one hand, if there are more INV crossbars and fewer VMM crossbars in the architecture, the latency of FP and BP graph can be longer because they are mainly composed of vector-matrix multiplication operations.
Otherwise, the WU graph, which is mainly composed of matrix inversion, needs more time to compute and may also reduce the throughput.
We conduct a design space exploration on this parameter to obtain the optimal architectural configuration.
As different neural networks have different layer sizes, the optimal architectural configuration can be different.
We use the averaged computational efficiency across all the benchmarks as a metric.
The result is shown in \figurename{}~\ref{fig:result_dse}.
The metrics increase when the proportion of VMM and INV crossbar grows.
The anomalous trends between the \#VMM~XB~/~\#INV~XB value 12 and 16 is becasue the buffer area grows non-uniformly when the buffer size grows.
When \#VMM~XB~/~\#INV~XB is larger than 32, the INV crossbar number is not large enough to arrange large NNs, e.g., VGG-19.
The best computational efficiency (722.1$\mathrm{GOPS/mm^2}$) is achieved when there are 28 VMM crossbars in each sub-tile, or $28\times 16=448$ VMM crossbars in each tile containing 16 INV crossbars.

\begin{table}
        \centering
\caption{\archname{}'s Area Breakdown (in $mm^2$).}
\vspace{-5pt}
        \label{tab:area_breakdown}
        \resizebox{\linewidth}{!}{
                \begin{tabular}{|c|c|c||c|c|c|}
                        \hline
                        Component & Spec                      & Area    & Component & Spec                      & Area    \\ \hline\hline
    \multicolumn{3}{|c||}{\textbf{VMM} XB} & \multicolumn{3}{c|}{\textbf{INV XB}}\\\hline
                        ADC       & \begin{tabular}{c}
                                Resolution: 8 \\
                                Freq: 1.2GSps \\
                                Num: 256      \\
\end{tabular} & 0.00236  & ADC       & \begin{tabular}{c}
                                Resolution: 8 \\
                                Freq: 1.2GSps \\
                                Num: 256      \\
\end{tabular} & 0.00236  \\ \hline
                        DAC       & \begin{tabular}{c}
                                Resolution: 4 \\
                                Num: 256      \\
\end{tabular} & 0.00068 & DAC       & \begin{tabular}{c}
                                Resolution: 4 \\
                                Num: 256      \\
\end{tabular} & 0.00068 \\ \hline
                        ReRAM     & \begin{tabular}{c}
                                Size: 256 \\
                                Num: 1    \\
                        \end{tabular} & 0.0001  & ReRAM     & \begin{tabular}{c}
                                Size: 256 \\
                                Num: 3    \\
                        \end{tabular} & 0.0003  \\ \hline
&                           &         & OpAmp     & Num: 512                  & 0.0128 \\ \hline
\textbf{Total}    & \textbf{Num: 28}                   & \textbf{0.0879}   & \textbf{Total}    & \textbf{Num: 1}                    & \textbf{0.0161}  \\ \hline \hline
\multicolumn{6}{|c|}{\textbf{Sub-Tile}}\\\hline
IR & Size: 4kB & 0.004 & OR & Size: 1kB & 0.002 \\\hline
Act       & Num: 1                    & 0.0006  & S+A       & Num: 29                    & 0.00174\\ \hline
                        Mul       & Num: 1                    & 0.0006  &           &                           &         \\ \hline
&                           &         & \textbf{Total}  & \textbf{Num: 16}                  & \textbf{1.80}   \\ \hline\hline
\multicolumn{6}{|c|}{\textbf{Tile}}\\\hline
eDRAM     & Size: 512kB               & 0.898   & Bus       &    Num: 928                       & 0.218  \\ \hline
&                           &         & \textbf{Total}      & \textbf{Num: 22}                   & \textbf{64.2}    \\ \hline\hline
Hyper Tr. & & 22.9 &&&\\\hline
& & &\textbf{Chip Total}&&\textbf{87.1}\\\hline \hline
                \end{tabular}
        }
\end{table}

With the results of design space exploration, the configuration of \archname{} is decided.
The eDRAM buffer of each tile is set to 512kB and the bus width is 256 bits.
With the inter-layer pipeline proposed in \cite{ISAAC}, this buffer size is enough to store the intermediate results.
Note that we clock the eDRAM at 1GHz while the crossbars are running at 10MHz, so that the eDRAM bandwidth should be enough for inner-tile data transfers.
The area of one \archname{} chip is 87.1$\mathrm{mm^2}$, including 22 tiles with 16 INV and 448 VMM crossbars per tile.
    This chip area is close to PipeLayer's \cite{pipelayer7920854}.
Area breakdown is presented in \tablename{}~\ref{tab:area_breakdown}.
    In the following experiments, we use 8 chips for both PipeLayer and \archname{} to train the benchmarks, so that the total area is close to the area of a single Tesla V100 GPU (815$\mathrm{mm^2}$).
    For smaller networks that can not occupy all the ReRAM crossbars in the 8 chips, we duplicate the matrices stored in ReRAM crossbars to speed up the training.

\subsection{Speedup and Energy Saving}
\label{evaluation:result:speed_energy}

\begin{figure}
    \centering
\includegraphics[width=0.9\linewidth]{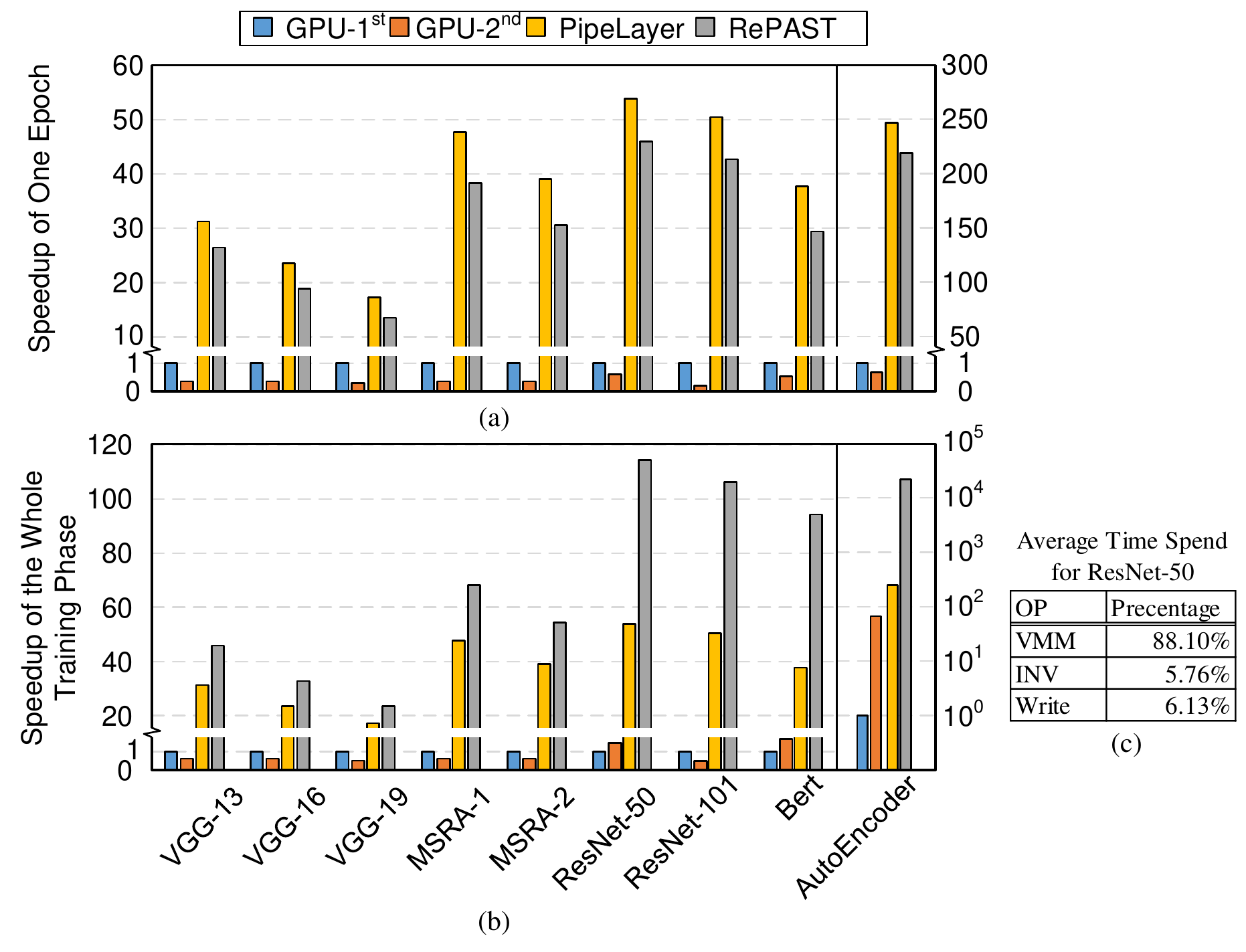}
\vspace{-5pt}
\caption{(a) (b) Speedup of \archname{} compared to GPU and PipeLayer over neural network benchmarks.  All the values are normalized to GPU-1$^{\textrm{st}}$. \revisionadd{Autoencoder use the secondary Y axis. (c) Average Time Spend on the operations when training ResNet-50}}
    \label{fig:result_speedup}
\vspace{-5pt}
\end{figure}

We compare the performance of \archname{}, GPU and PipeLayer in \figurename{}~\ref{fig:result_speedup}.
For each benchmark, we report the training time of the first-order training on GPU ($GPU-1^{st}$) and PipeLayer, and the second-order training on GPU ($GPU-2^{nd}$) and \archname{}.
The block size of the second-order training algorithm is set to $1024$. 
\figurename{}~\ref{fig:result_speedup} (a) depicts the training time of one epoch and \figurename{}~\ref{fig:result_speedup} (b) depicts the total training time.
For CNNs, we regard that the training is finished when the accuracy reaches within 0.5\% lower than the standard accuracy.
Compared to $GPU-2^{nd}$, \archname{} can achieve an average of $115.8\times$ reduction in training time across the benchmarks.
Compared to the first-order training on PipeLayer, \archname{} needs 21.5\% longer training time for each epoch on average.
However, thanks to the faster convergence rate in the second-order training which means much fewer epochs are needed for the entire training process, the total training time of \archname{} is $11.4\times$ shorter than PipeLayer.
In contrast, if only using GPUs, the second-order algorithm brings an average of 58.8\% more training time compared to
the first-order algorithm in GPU because the SOI inversion computation on GPU is too slow and may completely over-ride the benefit of fast convergence in the second-order algorithm.
For VGG-13/16/19, MSRA-1/2 and ResNet-101, the fast convergence rate cannot compensate for the inversion overhead on GPU.
For small-scale autoencoder, \archname{}'s per-epoch training time is 12.7\% longer than PipeLayer.
    Nevertheless, the convergence of second-order training on autoencoder is 109$\times$ faster than first-order training \cite{kfacpmlr-v37-martens15}.
This results in a 88.7$\times$ shorter training time compared to PipeLayer.

\revisionadd{
    Time Spend breakdown on the three operations when training ResNet-50 listed in \figurename{}~\ref{fig:result_speedup} (c).
    Note that vector-matrix multiplication and matrix inversion are actually computed in parallel with different crossbars.
    The time spend precentage is averaged number among the crossbars.
    Matrix inversion and writing only account for 11.9\% of crossbar's computing time because they only happens after every several batches.
}

\begin{figure}
    \centering
\includegraphics[width=0.7\linewidth]{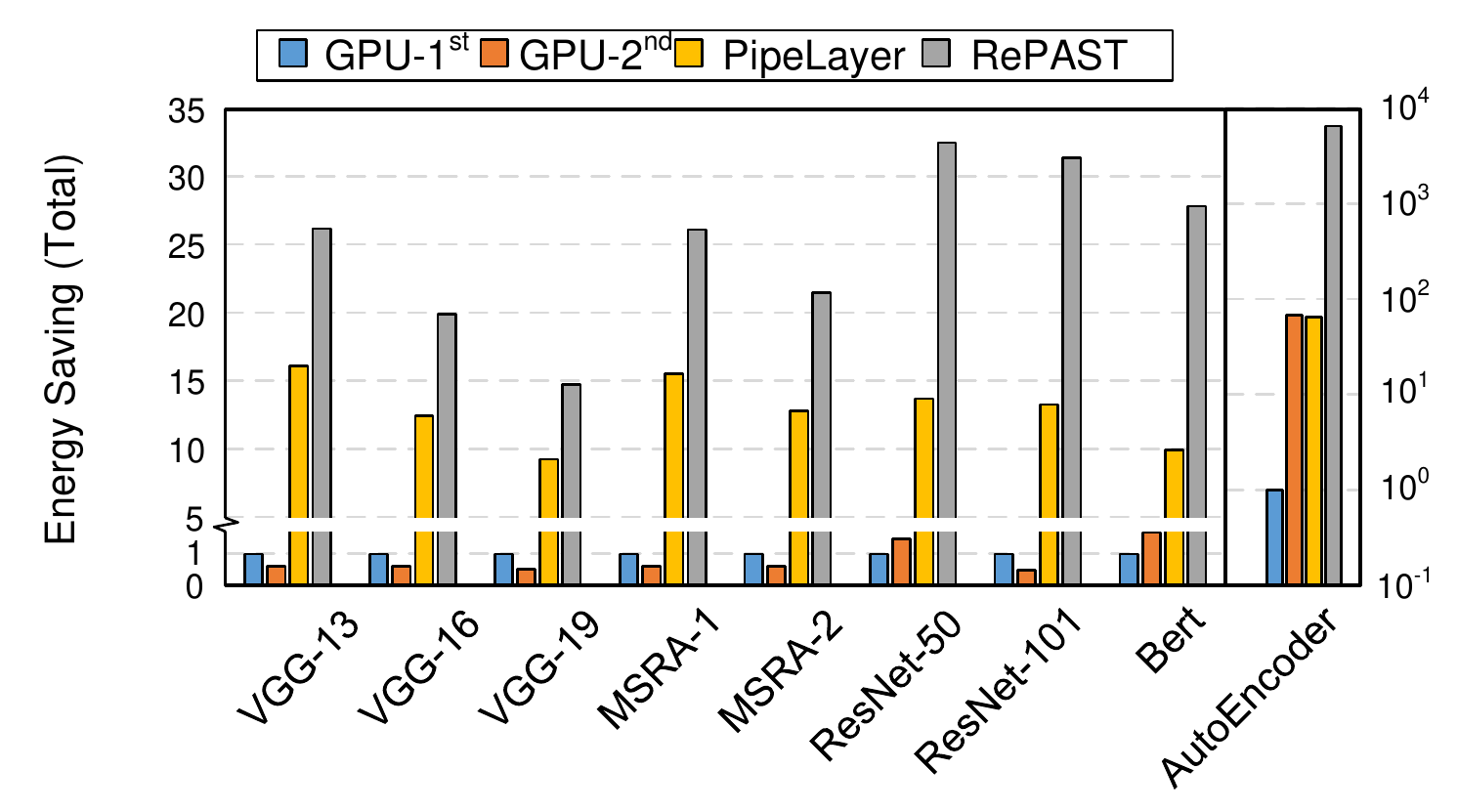}
\vspace{-10pt}
    \caption{Energy consumption of \archname{} compared to GPU and PipeLayer over neural network benchmarks. All the values are normalized to GPU-1$^{\textrm{st}}$.}
    \label{fig:result_energy}
\vspace{-10pt}
\end{figure}

The energy saving of \archname{} over GPU and PipeLayer is shown in \figurename{}~\ref{fig:result_energy}.
\archname{} can achieve an average energy saving of $41.9\times$ and $12.8\times$ compared to GPU and PipeLayer respectively.

\subsection{Write Number}

\revisionadd{
    \figurename{}~\ref{fig:res_block} (b) shows that 
    \archname{} can reduce 55.7\% write number on average during the whole training phase compared to PipeLayer.
    As with first-order training, weights on VMM crossbars are updated after each batch.
    SOI on INV crossbars are updated after several batches.
    Note that the matrices on INV crossbars keep unchanged during matrix inversion computing.
    Therefore, RePAST's write frequency is no higher than Pipelayers.
    Thanks to the fast convergance of second-order training, the iteration number of the whole training phase is reduced.
    This significantly reduce the writes and improve the endurance of \archname{}.
}

\subsection {Impact of the Mapping Scheme}

\begin{figure}
    \centering
\includegraphics[width=0.85\linewidth]{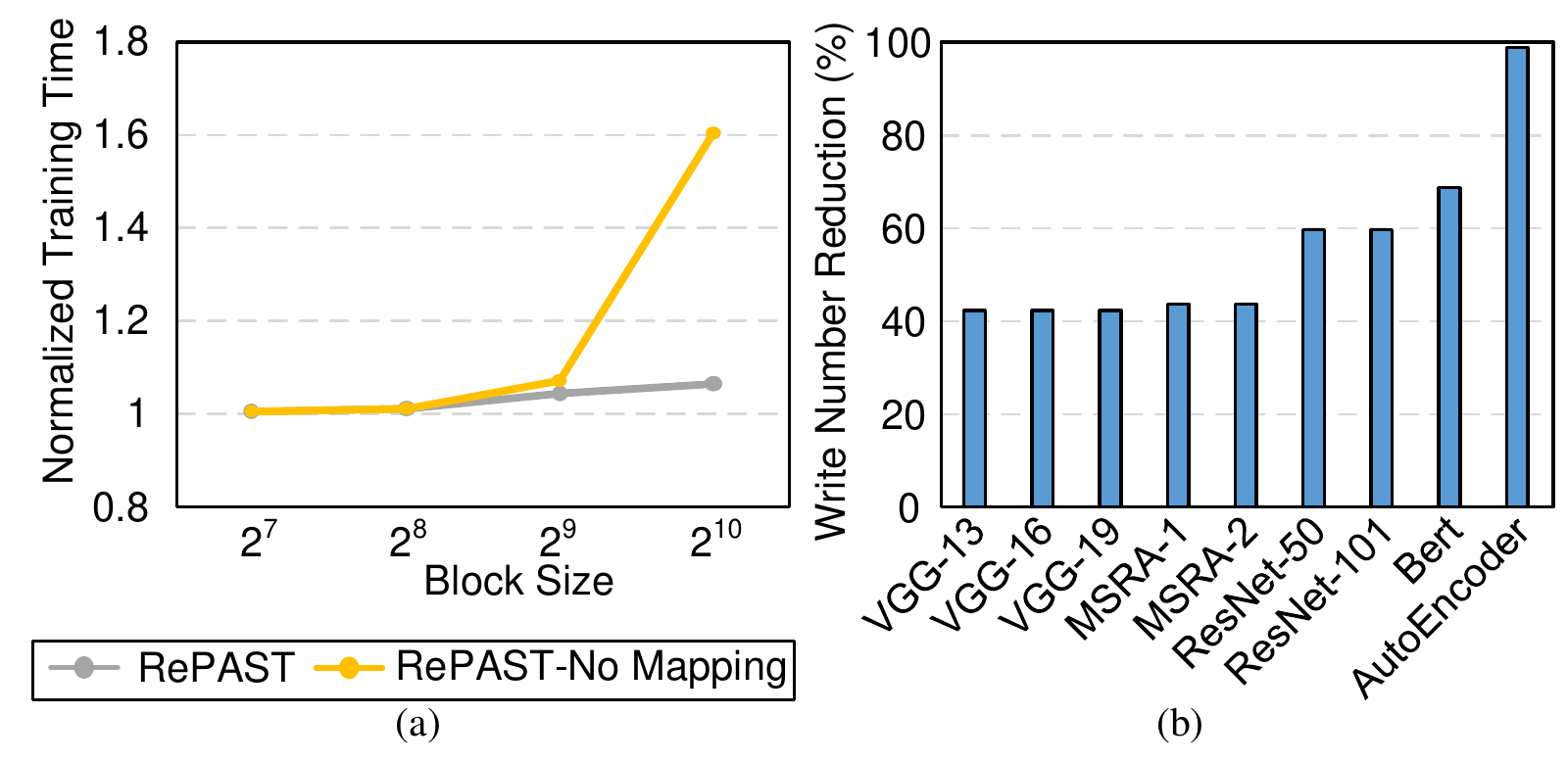}
\vspace{-10pt}
\caption{(a) Training time of \archname{} with/without mapping scheme over different block size. Normalized to the value on block size 128 of \archname{} with mapping scheme. \revisionadd{(b) Write number reduction of RePAST compared to PipeLayer.}}
    \label{fig:res_block}
    \vspace{-10pt}
\end{figure}

We evaluate the epoch training time of ResNet-50 on\archname{} with different block size and plot the results in \figurename{}~\ref{fig:res_block} (a) to study the impact of mapping schemes.
\archname{}-No Mapping denotes that the original dataflow is directly mapped to the \archname{} without using the mapping schemes proposed in Section~\ref{sec:mapping:pattern}.
When the block size is smaller than the crossbar size of $256$, the two series have the same training time.
This is because the small blocks in SOI matrices can not even occupy a single INV crossbar.
The first strategy of MM-INV pattern will always achieve the best saving.
When the block size becomes larger, the training latency of RePAST-No Mapping increases sharply.
This is because the crossbar occupation of SOI blocks increases squarely to the block size.
While with our proposed mapping schemes, the training latency increases more gently with a more flat slope.

We can illustrate that there is a maximum crossbar occupation of SOI matrices with the increasing of block sizes if we perform the mapping scheme for K-FAC algorithm.
Suppose the crossbar size is $s$.
The FIM for inversion $A$ is obtained by $A = a\cdot a^T$, and $A\in R^{c_{in}k^2\times c_{in}k^2}$, $a\in R^{c_{in}k^2\times hw}$.
The matrix $A$ is approximated by diagonal block matrix with block size $B$, with a total of $\left\lceil c_{in}k^2/B\right\rceil$ blocks.
Suppose the $i^\mathrm{th}$ block $A_i=A[iB:(i+1)B, iB:(i+1)B]$.
$A_i$ is computed by $a_i\cdot a_i^T$, where $a_i=a[iB:(i+1)B, :] \in R^{B\times hw}$.
For each block $A_i$, we can perform the mapping scheme, and the crossbar number is $\min\left(\left(\dfrac{B}{s}\right)^2, 2\dfrac{hw}{s}\cdot \dfrac{B}{s}\right)$.
When the block size $B$ is large ($B > 2hw$), this value will becomes $2hwB/s^2$.
For the whole matrix $A$, the crossbar number is $2hwB/s^2\cdot c_{in}k^2/B=2hwc_{in}k^2/s^2$, which is a constant independent to the block size $B$.
Therefore, the block size will not influence the resource occupation of \archname{} with our mapping schemes.
\revisionadd{
    Note that the conclusion is suitable for convolution layer of any size.
    We only need to instantiate the above parameters $c_in$, $hw$, $s$, $k$ into the real convolution layer's size.
}
This conclusion is meaningful because \archname{} is friendly to large block size and larger block size can take better advantage of the higher convergence rate in the second-order training algorithms.

\section{Related Work}
\label{sec:related}

\revisionadd{
    \textbf{ReRAM-based Accelerators.}
    \cite{ISAAC,prime10.1109/ISCA.2016.13,ReTransformer10.1145/3400302.3415640} design ReRAM-based accelerators for DNN reference.
    Some studies improve the energy efficiency of ReRAM-based DNN accelerator \cite{AtomLayer8465832,Timely,CASCADE,NeuralPIM}.
    \cite{pipelayer7920854,time8333741} extend the ReRAM-based accelerator to the first-order training for DNNs.
    The main problem of ReRAM-based accelerators for DNN training is limited endurance of ReRAM.
    \cite{longlivetime8465850} improve the endurance of ReRAM-based accelerator by swapping the rows during DNN training.
    \cite{AgingAware8714954} propose a weight-skewed training and aging-aware mapping.
    Another problem is the non-ideality in ReRAM's analog computation.
    \cite{Thermal10.1145/3195970.3196128,TheThermal10.1145/3400302.3415665} propose methods to tolerent the thermal noise.
    \cite{VAT10.1145/2744769.2744930,Friendly7926952} add the hardware non-ideality to the loss function and retrain the DNN.
    \cite{NoiseInjection10.1145/3316781.3317870} design a end-to-end framework to adapt the non-ideality.
    \cite{Unary9323041,alexnet10.1145/3065386} propose to use unary coding to reduce the impact of variance.
    \vspace{-10pt}
}

\revisionadd{
\textbf{Second-Order Training Algorithm.}
Newton method and Natural Gradient method are two traditional second-order optimizing methods.
The SOI's size is heavy for most applications.
The Quasi-Newton method \cite{quasinewtonnocedal2006numerical,lbfgs10.1145/279232.279236} uses the rank-1 matrices to approximate the Hessian matrix, but this size is still unacceptable for DNNs.
To further reduce the storage of Hessian matrix in Newton method, approximate Hessian \cite{spanhuang2020span, yao2021adahessian} and Hessian-free methods \cite{hessianfreekiros2013training, distributed8850515} are proposed.
\cite{kfacpmlr-v37-martens15,kfacrnnmartens2018kroneckerfactored,kfaccnnpmlr-v48-grosse16, shampooDBLP:journals/corr/abs-2002-09018} approximate the FIM in Natural Gradient method.
SOI change slowly during training. Therefore, \cite{spngdosawa2020scalable, thorChen_Gao_Liu_Wang_Ni_Zhang_Chen_Ding_Huang_Wang_Wang_Yu_Zhao_Xu_2021} use the outdated SOI to reduce the update frequency.
}
\section{Conclusion}
\label{sec:conclusion}
In this work, we propose \archname{}, a ReRAM-based PIM accelerator for the second-order training of DNN.
We propose a novel high-precision matrix-inversion algorithm and implementation with normal VMM and INV crossbars.
The INV crossbar in \archname{} is connected with analog wires and can support a fused operation of low-precision matrix inversion and its immediately leading low-precision matrix multiplication.
Based on this feature, we propose a mapping scheme for MM-INV and successive MM/INV pattern in the algorithms.
Results show that \archname{} can achieve an an average of 115.8$\times$/11.4$\times$ speedup and 41.9$\times$/12.8$\times$ energy saving compared to a GPU counterpart and PipeLayer on large-scale DNNs.
Moreover, the \archname{} is not sensitive to the second-order information and can better take advantage of the high convergence rate of the second-order training algorithms.

\nocite{nocite110.1145/3400302.3415666,nocite210.1145/3466752.3480071}


\bibliographystyle{IEEEtranS}
\bibliography{refs}

\end{document}